\documentclass[a4paper,11pt]{scrartcl}
\usepackage{graphicx}
\usepackage[footnotesep=12mm]{geometry}
\usepackage{amsmath}
\usepackage{amssymb}

\begin{document}
  \title{ Coupling of cytoplasm and \\
          adhesion dynamics determines \\ 
          cell polarization and locomotion}
  \makeatletter
  \renewcommand{\@fnsymbol}[1]{\@arabic{#1}}
  \makeatother
  \author{
    Wolfgang Alt\thanks{ corresponding author, eMail 
                         \texttt{wolf.alt@uni-bonn.de} 
          } $^,$\thanks{ Universit\"at Bonn, Theoretische Biologie, 
                         Kirschallee 1-3, 53115 Bonn, Germany } \and
    Martin Bock\footnotemark[2] \and
    Christoph M\"{o}hl\thanks{ Forschungszentrum J\"ulich, 
              Institut f\"ur Bio- und Nanosysteme (IBN), \newline 
              Institut 4: Biomechanik (IBN4), 52425 J\"ulich, Germany }
  }
  \maketitle
  \thispagestyle{empty}
  \begin{abstract}
    Observation of epidermal cells or cell fragments on flat adhesive
    substrates has revealed two distinct morphological and functional states:
    A non-migrating symmetric "unpolarized state" and a migrating asymmetric
    "polarized state". They are characterized by different spatial
    distributions and dynamics of important molecular components as F-actin and
    myosin-II within the cytoplasm, and integrin receptors at the plasma
    membrane contacting the substratum, thereby inducing so-called focal
    adhesion complexes. So far, mathematical models have reduced this
    phenomenon either to gradients in regulatory control and signaling
    molecules or to different mechanics of the polymerizing and contracting
    actin filament system in different regions of the cell edge.

    Here we offer an alternative self-organizational model in order to
    reproduce and explain the emergence of both functional states for a certain
    range of dynamical and kinetic model parameters. We apply an extended
    version of a two-phase, highly viscous cytoplasmic flow model with variable
    force balance equations at the moving edge, coupled to a 4-state
    reaction-diffusion-transport system for the bound and unbound integrin
    receptors beneath the spreading cell or cell fragment. In particular, we
    use simulations of a simplified 1-D model for a cell fragment of fixed
    extension to demonstrate characteristic changes in the concentration
    profiles for actin, myosin and doubly bound integrin, as they occur during
    transition from the symmetric stationary state to the polarized migrating
    state. In the latter case the substratum experiences a low magnitude
    "pulling force" within the larger front region, and an opposing high
    magnitude "disruptive force" at the shorter rear region. Moreover,
    simulations of the corresponding 2-D model with free boundary show
    characteristic undulating protrusions and retractions of the cell
    (fragment) edge, with local accumulation of doubly bound adhesion receptors
    behind it, combined with a modulated retrograde F-actin flow. Finally, for
    a stationary model cell (fragment) of symmetric round shape, larger
    fluctuations in the circumferential protrusion activity and adhesion
    kinetics can break the radial symmetry and induce a gradual polarization of
    shape and concentration profiles, leading to continuous migration in
    direction of the leading front.

    The aim of the chapter is to show how relatively simple laws for the
    small-scale mechanics and kinetics of participating molecules, responsible
    for the energy consuming steps as filament polymerization, pushing and
    sliding, binding and pulling on adhesion sites, can be combined into a
    non-linearly coupled system of hyperbolic, parabolic and elliptic
    differential equations that reproduce the emergent behavior of
    polarization and migration on the large-scale cell level.
  \end{abstract}

\tableofcontents

\graphicspath{{figures/}}

\section{Biology of cell polarization and migration}\label{sec:biology}

Cell polarization and migration plays a central role in the
development and maintenance of tissues in multicellular organisms.
During ontogenesis, new tissues are formed by the coordinated
division and locomotion of single cells. The polarization of a cell
not only defines the direction of migration \cite{Jiang05} but also
the cell division axis \cite{Thery05} and thus the three-dimensional
structure of tissues, organs and finally the whole organism.

\subsection{Asymmetry of actin polymerization and
            substrate adhesion}\label{sec:asym}

The coordinated development and release of focal adhesions (FAs) is
a basic requirement for directed cell migration. Migrating cells
feature pronounced adhesion dynamics and a structural polarity with
a clearly distinguishable frontal and rear area. Actin
polymerization predominates at the cell front resulting in a
protruding lamella in direction of migration. New focal adhesions
composed of clustered protein complexes develop at the lower
membrane of the lamella near the leading edge and couple the
F-actin-network mechanically to extracellular matrix proteins.
Simultaneously, the matured FAs residing at the opposed trailing
edge are dissolved while myosin driven contraction of F-actin moves
the cell body forward
\cite{burridge:focalassembly,lauffenburger:cellmigration}.

Arising from the process described above, the polarity of the cell
can be referred to two structural asymmetries, which are key
requirements for effective cell migration: asymmetry of actin
polymerization and asymmetry of adhesion strength. The growth of
actin filaments has to predominate at the cell front for pushing the
leading edge in direction of migration
\cite{verkhovsky:polarization,Sma02}. To move the cell body forward,
it has to be released from the substrate during contraction by
dissolving the FAs at the back (rear release) while the FAs at the
cell front have to remain stable to provide a mechanical attachment
for the contractile machinery pulling the cell body. In absence of
rear release, traction forces could be dominated by adhesion forces
and the cell would get stuck \cite{ridley:integrating}. In this
regard, the spatial distribution of adhesion strength and actin
polymerization defines the direction of migration and could be
specifically regulated due to directed cell movement.

\subsection{Flow of actin filaments and myosin gradient}

The mechanisms underlying the structural polarity of migrating cells
are still under discussion, particularly what concerns an intrinsic
directionality of the cytoplasm. For example, the finding that
asymmetric adhesive structures define polarization of a touching
cell \cite{Jiang05} suggests that directional flow of the actin
cytoskeleton is involved in the process of cell polarization.
Moreover, recent studies on fish keratocytes have revealed that
polarization occurs spontaneously and is accompanied by a
reorganization of the actin cytoskeleton, which finally leads to
cell locomotion \cite{pmid17893245}. In these experiments,
unpolarized solitary cells feature a circular shape with a radially
symmetric actin distribution. Although the cells do not move,
transient protrusions and retractions appear at the cell edge, and
actin flows centripetally with a decreasing flow gradient from the
cell edge to the center. In this apparently unstable state
spontaneous symmetry breaking results in a faster inward flow and in
an increased concentration of actin at the rear region of the cell.
On the opposed front edge, the reduced inward actin flow causes
protrusion and the development of a lamellipodium. In this polarized
state, the cell starts to migrate while attaining a more or less
constant shape.

A similar behavior was observed with cell fragments extracted from
the lamella of fish keratocytes \cite{verkhovsky:polarization},
\begin{figure}
  \centering
  \includegraphics[width=0.70\textwidth]{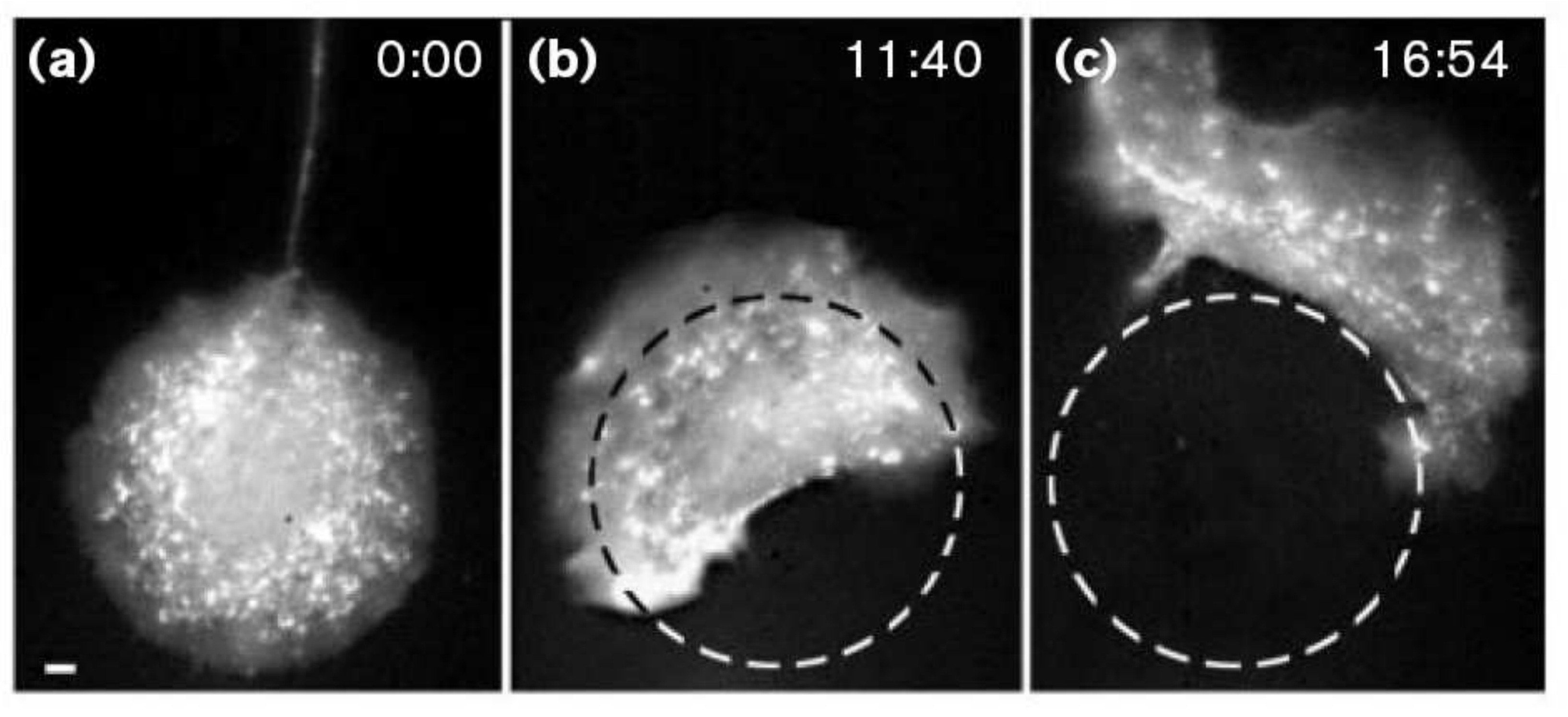}
  \caption{Cell fragment experiments
           where the mechanical stress (release) induces a transition
           from a circular non-migrating state (a) to a polarized
           migrating state (b,c). 
           The cell fragment was labelled with fluorescent myosin II.
           Time scale is min:sec and length scale of bar is 
           $2 \, \mu \mathrm m$
           (from \cite{verkhovsky:polarization}).
  }\label{fig:expFragmente}
\end{figure}
see Fig.~\ref{fig:expFragmente}. 
Since these fragments were lacking
most cell organelles and the microtubule system, they were mainly
limited to the actin-myosin machinery. They appeared either in an
unpolarized, non-migrating state with homogeneous actin-myosin
distribution or a polarized, migrating state with a rising myosin
concentration from front to rear. Interestingly, polarization of
fragments could be induced by mechanical stimulation, e.g.~by shear
flow or stress release leading to a transition from the stationary
to the migrating state.

Cell polarization is discussed to result from the local activation
of GTPases by the rho-family
\cite{Tur00,ridley:integrating,mogilner:motility}. These proteins
are known to regulate actin polymerization, myosin activity and FA
assembly, whereas it remains unclear how the distribution of the
GTPases is controlled. Recent experiments have revealed that by
inhibition of myosin activating signalling pathways the cell's
ability to polarize is reduced \cite{pmid17893245}. However, the
observations of cells or cell fragments to polarize either
spontaneously or due to mechanical perturbations, raise doubts that
biochemical signalling is the primary reason for inducing and
maintaining polarity.
Experiments focussing on cell mechanics rather show that direction
and strength of locomotion forces are inherently
connected with the retrograde F-actin flow \cite{Pon04,Gardel08}.
This suggests that the mechanical action of myosin, namely to induce
cytoplasmic contraction, flow and force transduction at adhesive
sites, plays a key role in explaining the ubiquitously observed
phenomena of cell polarization and locomotion.

\section{Previous models of cytoplasm and adhesion dynamics}

The first detailed mathematical model coupling cytoskeleton and
adhesion dynamics was developed by Lauffenburger and coworkers
almost two decades ago \cite{DiMillaLauffenburger}. Here, the
contractile actin-myosin network and its interaction with bound
adhesion sites is represented as a mechanical system of connected
viscoelastic units of generalized Kelvin-Voigt type, which
constitute the (three) inner segments as well as a front segment
(lamellipod) and a tail segment (uropod) for a rectangular model
cell of fixed width and length. A coupled system of
reaction-diffusion equations for free and substrate bound integrins
or adhesion receptors on the \emph{dorsal} (lower) and
\emph{ventral} (upper) part of the model cell is solved under pseudo
steady-state assumptions. This dynamically provides the number of
adhesion bonds in both end segments, lamellipod and uropod, whereby
their rupture (dissociation) kinetics exponentially increases with
force load onto a bound adhesion site. The resulting non-linear
dynamics for the local (forward or backward) displacement of each
viscoelastic unit produces a persistent forward translocation of the
whole model cell. However, this is achieved
only if a front-tail asymmetry is presupposed, either 
by exposing more free adhesion receptors, or 
by assuming higher adhesion bond affinity at the front compared to
the tail. The authors present a series of results on how the
simulated cell migration speed depends on various model parameters
as, for example, cytoskeletal contractility or adhesion strength. A
later variant of this model presents a more explicit study of
adhesion bond disruption kinetics at the rear of the cell and
already uses a four-state model for integrin binding to the
cytoskeleton and to the substratum \cite{Lauffenburger99}.

In recent years a series of more elaborate models
have been developed, accounting for details of
the meanwhile discovered molecular regulation mechanisms
for the chemical and mechanical processes,
particularly at the free boundary of a moving cell.
One model type is based on spatially discrete algorithms (cellular
Pott's model) using the definition of local energies to determine
the protrusion and retraction of boundary elements
via the stochastic Metropolis rule,
e.g.~\cite{Maree,NishimuraS2,NishimuraUeda}.

Another class of mechanical cell models takes into account branching
and anisotropy of cytoskeletal actin filaments at various
times and in various regions of the cell, see
e.g.~\cite{Zhu88,gholami:oscillations08,Huber08}. Besides explicit
constructions of an elastically cross-linked network in the leading
edge \cite{oelzschmeiser,KoestlerSmall} biophysical models have been
developed to capture different dynamics of Arp2/3-induced branching
and myosin-induced contraction of F-actin networks by Mogilner and
coworkers
\cite{Mog03,GrimmMogilner,mogilner:polarization}. The last article
is particularly devoted to explain the mentioned polarization
experiments with cell fragments (see Fig.~\ref{fig:expFragmente}) by
assuming a different actin-myosin network organization at the rear
in comparison to the front region.

Based on the `reactive flow' model by Dembo and coworkers
\cite{DemboHarlow,dembo1989,altdembo}, the most elaborate
model extension has been presented by Oliver et al. \cite{Oli05}:
they use full \mbox{3-D} two-phase flow equations with free ventral
boundary and with two additional rapidly diffusing messenger
concentrations that regulate actin network contractility and
(de-)polymerization. Moreover, cell adhesion is modeled by a
Navier-slip boundary condition at the substratum, in which only
constant adhesive properties are taken into account. The analysis of
this complex model, being performed in the thin film limit, is
restricted to linear-stability arguments for ruffle generation and
to local expansion analyses at the moving tip; there,
phenomenological equations for boundary mass fluxes are considered
without specifying the types of molecular mechanisms for tip
protrusion. Finally, quantitative estimations for the pseudopod
protrusion and cell translocation speed under various sub-limit
assumptions are given, which turn out to be consistent with observed
values, particularly for osteoblasts, though no numerical
simulations are given that would re-insure the analytic results.
A similar thin-film approximation of the \mbox{2-D} equations under
incompressibility assumptions for a `viscous polar gel' has been
used in \cite{Kruse06} to derive explicit expressions for the
advancing speed of a cell lamella, but again by predefining its
polarity.

Except the last one, all so-far mentioned models do not explicitly
quantify the varying force field, which is applied by the
cytoskeleton onto the substrate covered by a migrating cell and
which has been approximately reconstructed by different inverse
methods from experimental assays of cells, \mbox{e.g.} moving on
flexible substrata \cite{wangdembo,Lo00,schwarzbalaban,merkel:finitethickness,ambrosi:tractionPatterns}.
A first model implementing force transduction to the substrate has
been proposed by Gracheva and Othmer \cite{grachevaothmer} by
specifying a spatially one-dimensional system of viscoelastic
equations for cytoskeleton dynamics, whose polymerization,
contractility and adhesive binding is regulated by signalling
molecules. However, they make a pseudo-steady-state assumption for
the binding kinetics of myosin polymers to actin filaments, and of
transmembrane integrin proteins to the substratum. Moreover, they
impose artificially defined gradients from tail to front of certain
regulatory proteins in order to stimulate polarized cell
translocation.

Recently, adhesion kinetics has also been implemented into an
extended cytoplasm flow model \cite{AltTranquillo95,AltTranquillo97}
describing the F-actin dynamics in an annular domain and its
coupling to lamellipodial protrusions and retractions
\cite{StephanouTracqui}. Force dependent maturation of FA complexes
and active polymerization of F-actin enable the simulation model to
reproduce characteristics of fibroblast shape deformation and
translocation.

In an earlier publication we have presented another extension of the
basic two-phase flow model for the 2-dimensional viscous cytoplasm
dynamics \cite{Alt03} by coupling the constitutive
hyperbolic-elliptic equations  to a system of four
reaction-diffusion-transport equations for the integrins beneath the
cell or cell fragment \cite{kuuselaalt}. Here we propose a
generalized continuum model, in which we couple cytoplasm and
adhesion dynamics with mechanical tension and transport of the
plasma membrane, in order to reproduce spontaneous and induced cell
polarization leading to migration, with assembly of adhesion sites
at the cell front and adhesion release at the cell's rear end.
Moreover, the model exhibits typical features of migrating cells as
protrusion/retraction cycles, rearward actin flow, pulling forces
at the front and a concentration of disruptive forces at the rear.
In the model, the cytoplasm is described as a viscous and
contractile fluid of polymers representing the actin cytoskeleton
interpenetrated by an aqueous phase. This actin filament network is
preferentially assembled at the cell edge, and can be contracted by
cross-linking with diffusing myosin oligomers. The moving actin
filaments then couple to transmembrane adhesion proteins that are
freely diffusing in the membrane or bound to the substrate on the
extracellular domain. Thus, the cytoskeleton mechanically connects
to the substrate through dynamic binding processes which results in
force transduction and finally cell locomotion.

Throughout our model presentation we rely on continuum descriptions,
in which macroscopic mass and momentum laws are combined with
mesoscopic submodels for fast molecular kinetics due to adequate
pseudo-steady-state assumptions.

\section[Multiply coupled reaction-diffusion-flow model]{Two-phase
flow model for cytoplasm coupled to reaction, diffusion and
transport for myosin-II and integrin proteins}

For simplicity, we restrict our model derivation and analysis to a
flat two-dimensional geometry, so that cells or cell fragments are
assumed to be homogeneously spread on the substrate without
considerable change of cell height. Thereby, three-dimensional
effects around the cell nucleus (e.g.~due to cell rolling) or along
the ventral (upper) plasma membrane are neglected. In order to
reproduce the main biophysical mechanisms and biochemical processes
that enable a cell to polarize and translocate on an adhesive
substratum, we nevertheless distinguish between the cytoplasm and
the exterior plasma membrane: On the one hand, F-actin assembly,
myosin kinetics and viscous network flow take place within the
cytoplasmic interior of the cell. On the other hand, integrin
binding to substrate or cytoskeleton and its transport or diffusion
are confined to the dorsal (lower) plasma membrane, where forces are
transduced to or generated at the cell periphery, leading to tip
protrusion or retraction. Thus, the moving cell (fragment) is simply
represented by a time-dependent connected domain $\Omega(t) \subset
\mathbb R^2$, over which the cytoplasmic volume extends with fixed
constant height, and any ruffles or blebs on top of the cell are
neglected. We rather assume that the cell dynamics is completely
determined by a flat cytoskeleton sheet of F-actin network,
\begin{figure}
  \centering
  \includegraphics[width=0.7\textwidth]{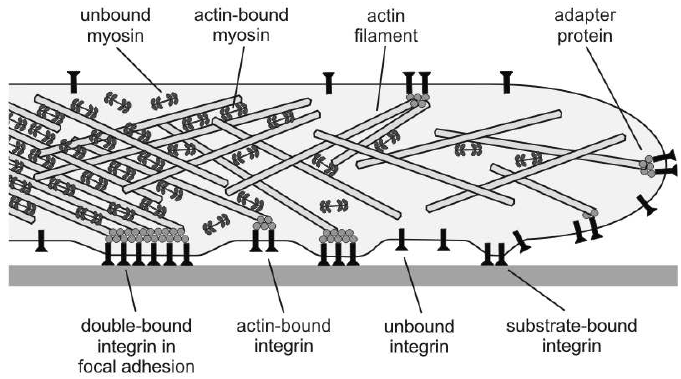}
  \caption{Schematic longitudinal section through a
           cell fragment (or a cell lamella) of constant height
           with involved proteins. For further explanation and model
           derivation see the following sections.
  }\label{fig:lamellipodium}
\end{figure}
see Fig.~\ref{fig:lamellipodium}.
This cross-linked filament phase with volume fraction
$\theta(t,\mathbf x), \mathbf x \in \Omega(t)$
is of constant thickness and connected
to the upper and lower plasma membrane
in such a way that also the cytosol,
i.e.~the solvent phase with volume fraction $(1-\theta(t,\mathbf x))$,
is confined to the same volume space.

By suitable model simplifications, we aim to capture the
self-organizational power of the cytoplasm as a two-phase fluid
coupled with the reaction, transport and diffusion of a series of
chemical ingredients. We explicitly model the kinetics and dynamics
of only those F-actin associated proteins that induce the main
bio-mechanical processes of force generation and transduction to the
substratum, namely myosin oligomers and transmembrane integrin
proteins. All other regulatory proteins such as e.g.~Rac, Rho or the
branching protein Arp2/3 or smaller substrate molecules,
e.g.~monomeric G-actin, are assumed to rapidly diffuse within the
cytosol under fast regeneration, so that they attain constant
reservoir concentrations serving as parameters in the model.

\subsection{Mass balance and flow equations}\label{sec:massbalanceeqs}

In the chosen two-dimensional continuum description, cell
deformation and translocation with respect to the fixed substratum
is represented by movement of the cell edge $\Gamma(t) =
\partial\Omega(t)$ which, however, is induced by three distinguished
mass flows with potentially different transport velocity fields on
$\overline{\Omega}$, namely the mean velocities $\mathbf v$ of the
F-actin cytoskeleton ($\theta$) and $\mathbf w$ of the aqueous
cytosol ($1-\theta$) plus the transport velocity $\mathbf u$ of the
lower dorsal plasma membrane. Thereby both $\mathbf v$ and $\mathbf
w$ are averaged over the constant cell height. For the membrane area
flux $\mathbf u$ we make the simplifying assumption that the lipid
bilayer between the flat substratum and an adhering cell is always
stretched out to its maximal extension with constant area
concentration of the lipid-protein mixture, thus constituting an
incompressible two-dimensional Newtonian fluid. Then, freely
diffusing transmembrane integrin proteins are additionally
transported by the membrane velocity $\mathbf u$, myosin oligomers
by the cytosol velocity $\mathbf w$, and F-actin bound myosin-II
motor molecules by the cytoskeleton velocity $\mathbf v$. The
detailed mass balance equations are discussed in the following
paragraphs.

\subsubsection{Mass conservation for the two phases cytoskeleton and cytosol.}
Cytoskeleton and cytosol in the flat two-dimensional geometry can
experience counteracting flows due to local contraction, assembly or
relaxation of the F-actin network, while the bulk cytoplasmic fluid
with constant volume fraction $1$ can be assumed to be
incompressible, at least in the range of occurring pressure
differences $\lesssim \mathrm{kPa}$. Compare the analogous situation
of a water-filled sponge that is internally condensing without
changing its shape. Then, because of the fixed height assumption,
the total two-dimensional volume flux $\mathbf W = \theta\mathbf v +
(1-\theta)\mathbf w$ is divergence free at any time $t$, yielding
the first local mass conservation law
\begin{equation}
\label{eq:DivergenceFree}
\nabla \cdot \Bigl( \theta\mathbf v + (1-\theta)\mathbf w \Bigr) = 0.
\end{equation}
Therefore, also the total cell volume (i.e.~the 2-dimensional cell
area) is conserved over time, so that the `bulk' fluid moves
together with the free boundary $\Gamma(t)$, meaning that the total
volume flux $\mathbf W$ has to fulfill the natural free boundary condition
\begin{equation}
\label{eq:NormalFlow} \nu_\Gamma\cdot \mathbf W = \dot\Gamma
\end{equation}
with $\nu_\Gamma$ denoting the exterior normal of $\Gamma$ and
$\dot\Gamma$ quantifying the normal speed of the cell edge.

In addition to possible convection, the F-actin network can locally
be assembled by filament polymerization from the pool of monomeric
G-actin within the cytosol and it can be disassembled by the reverse
process of severing or depolymerization. This mass exchange between
the two phases is given by a second conservation law, which for the
filament volume fraction $\theta$ can be written as
\begin{equation}\label{eq:MassBalance}
\partial_t \theta + \nabla\cdot(\theta \mathbf v) = R(\theta)
\end{equation}
with a \emph{net assembly rate} $R(\theta)$ to be modeled. According
to our simplifying assumption, the dependence on G-actin, Arp2/3 and
possible regulatory proteins enters only via constant parameters:
\begin{equation}
\label{eq:Assembly} R (\theta) =
\left( k_\mathrm{on} a_g - k_\mathrm{off} \right) B(\theta) - r \theta
\end{equation}
with concentration of globular actin $a_g$, polymerization rate
$k_\mathrm{on}$ and depolymerization rate $k_\mathrm{off}$ at the
filaments' plus (barbed) ends, as well as another lumped disassembly
rate $r$, see e.g.~\cite{Mog02}. Due to relatively fast nucleation
and capping of actin filaments, the \emph{relative number} $B$
\emph{of barbed ends} is assumed to stay in pseudo-equilibrium with
the local F-actin concentration $a = \theta \cdot a_\mathrm{max}$,
where we suppose a maximal condensation of actin filaments in the
order of $a_\mathrm{max} = 800 \, \mu \mathrm M$. Following
\cite{Mog02} we write
\begin{equation} \label{eq:BarbedEnds}
B (\theta) =
    \frac{1}{\omega} \left(
      \varepsilon + \nu_0 \frac{ \theta }{ K_a / a_\mathrm{max} + \theta}
    \right)
\end{equation}
with capping rate $\omega$, a basic nucleation rate $\varepsilon$
and an induced branching rate $\nu_0 = \nu_a \mathrm{Arp0}$,
proportional to the concentration $\mathrm{Arp0}$ $[\mu \mathrm{M}]$
of activated Arp2/3, together with a half-saturation concentration
$K_a$ for its primary actin binding
site.

\subsubsection{Reaction-transport-diffusion equations for myosin oligomers.}
Myosin-II oligomers are the most important actin binding proteins
that are responsible for the generation of contractile forces within
cross-linked F-actin networks. Thus, their spatial distribution
within a polarizing or moving cell plays a key role for the
cytoplasm dynamics. In a most simple way we only distinguish between
freely diffusing myosin oligomers ($m_f$) and those that are bound
to cytoskeletal actin filaments ($m_b$) and, therefore, are
convected with velocity $\mathbf v$. Since free myosin tetramers
first have to attach to the actin network at one binding site, we
consider this bimolecular reaction ($m_f$ with $a$) as the rate
limiting step. Correspondingly we assume, that the relatively faster
processes of double binding ($m_b$ with $a$) and power stroke
formation are in a pseudo-steady state. Thereby the contractile
stress $\psi = \psi_0 m_b \theta$ with $\theta = a / a_\mathrm{max}
$ in the network is determined, see equation (\ref{eq:Contraction})
in the following section. Finally, we suppose a constant diffusivity
$D_m$ for free myosin oligomers, embedded into the cytosol flow
$\mathbf w$. Under these assumptions the local mass balance
equations for the corresponding myosin concentrations are:
\begin{align}
  \partial_t m_f &= \nabla \cdot \Bigl( D_m \nabla m_f -  m_f \mathbf w \Bigr)
                    - \alpha_m\cdot a \cdot m_f + \delta_m(a) \cdot m_b
                 \label{eq:dtmf} \\
  \partial_t m_b &= - \nabla \cdot \bigl( m_b \mathbf v \bigr)
                    + \alpha_m\cdot a \cdot m_f
                    - \delta_m(a) \cdot m_b
                 \label{eq:dtmb}.
\end{align}
For the association rate $\alpha_m$ we assume a constant
parameter, whereas the dissociation rate is supposed to increase
quadratically with increasing network concentration $a$ because of
steric inhibition or competition for binding sites:
\begin{equation}  \label{eq:delta-m}
  \delta_m (a) =
      \delta_m^0 \biggl( 1 + \frac{a^2}{a_\mathrm{opt}^2} \biggr) =
      \delta_m^0 \biggl( 1 + \frac{\theta^2}{\theta_\mathrm{opt}^2} \biggr),
\end{equation}
with an optimal actin network concentration $a_\mathrm{opt} = a_\mathrm{max}
\theta_\mathrm{opt}$. Moreover, in the case of fast diffusion and no transport,
the constant equilibrium concentrations $m_f^*$ and $m_b^*$ satisfy
the equality
\begin{equation}
\label{eq:MyosinStar} m_b^* = \frac{ \alpha_m a } {\delta_m(a) } m_f^*,
\end{equation}
so that the resulting contractile stress $\psi(\theta)$
becomes optimal for $\theta = \theta_\mathrm{opt}$,
Fig.~\ref{fig:assembly} below.

\subsubsection{Mass conserving flow of the dorsal plasma membrane.}
As mentioned before, we will assume that the dorsal plasma membrane
beneath an adhering cell (fragment) is stretched and that small
fluctuations can be neglected. On the contrary top side, the ventral
plasma-membrane usually shows extensive wrinkles or folds, which
most probably are induced by the contractile action of the
cytoskeleton itself, indicating that the plasma membrane tip at the
cell edge stays under a positive tension $\tau^\Gamma$. Before
discussing the corresponding dynamics of the dorsal membrane,
we state the two extreme possibilities, namely whether the membrane
moves together with the cell or not:
\begin{enumerate}
\item \emph{Membrane sticking to substrate:}
  During cell translocation over the substratum the
  whole dorsal membrane stays fixed and no slip may occur
  due to relatively strong interaction forces with the substratum,
  e.g.~via the glykocalix, so that $\mathbf u\equiv0$.\\
\item \emph{Membrane sticking to cell edges:}  The dorsal membrane is
  pulled along the substratum due to cytoplasm protrusions at the
  leading front, but with no slip around the lamellar tips
  due to strong membrane curvature. In this way the membrane area
  flux satisfies the mass conservation law together with a boundary
  condition analogous to (\ref{eq:NormalFlow}), namely
  \begin{equation}\label{eq:MembraneFlow}
  \nabla \cdot \mathbf u = 0 \; \; \text{on} \; \Omega(t) \qquad \qquad
  \nu_\Gamma \cdot \mathbf u = \dot\Gamma \; \; \text{on} \; \Gamma(t).
  \end{equation}
\end{enumerate}
In the latter case, the membrane can slip over the substratum,
thereby experiencing a finite frictional drag force, see
(\ref{eq:Tension_u}) below, whereas in the first case this force is
infinitely large to suppress slipping -- but then the membrane has
to slip around the tips at moving cell edges. Clearly, a certain
mixture of both possibilities is physically realizable but not
considered here.

Moreover, frictional drag onto the dorsal membrane can also occur at
its cytoplasmic side. Whereas the relative flow of cytosol will have
negligible effects, the horizontal F-actin flow develops a vertical
flow profile. This profile depends on the amount of `Navier-slip' in
a cortical layer on top of the dorsal membrane and emerges due to
frictional forces between actin filaments and membrane proteins. In
analogy to the more explicit thin film approximation \cite{Oli05} we
only consider the averaged profile velocity $\mathbf v$. In this way
we can introduce a simplifying \emph{cortical slip parameter}, $0 <
\kappa < 1$, so that the effective relative velocity between
cytoskeleton and dorsal membrane is reduced to $\kappa (\mathbf v -
\mathbf u)$. Thus, the resulting \emph{effective velocity of
cortical F-actin} in a thin layer above dorsal membrane and
substratum is given by
\begin{equation}
\label{eq:CorticalFlow} \mathbf v_c = \kappa \mathbf v + (1-\kappa)\mathbf u.
\end{equation}
Here the factor $\kappa$, which clearly depends on the viscous shear
properties of the cytoskeleton, should be larger than $0.5$ to
reflect observations of a generally slippy behavior (see
e.g.~\cite{Jia03,Gardel08}). Moreover, we suppose that the vertical
profile, thus $\mathbf v_c$, is not changed if some of the cortical
actin filaments are (transiently) bound to integrins in the dorsal
membrane, see Fig.~\ref{fig:integrins} and the following paragraph:
Either those bound integrins or integrin complexes are passively
pulled through the lipid-protein bilayer with relative velocity
$\mathbf v_c-\mathbf u$ or, in case of substrate-fixed adhesion
bonds, with relative velocity $-\mathbf u$. The whole F-actin
network above such a bond is slowed down by a certain local
frictional force per adhesion site,
\begin{equation}
\label{eq:CorticalFriction} \mathbf F_c = \Phi_0 \theta  \mathbf v_c\, ,
\end{equation}
entering into the corresponding force balance law, see
(\ref{eq:Phthcsa}) below. In any case, the frictional drag onto the
dorsal membrane, induced by the relative motion of singular integrin
complexes, will be neglected in our model.

\subsubsection{Reaction-transport-diffusion equations for membrane integrins.}
The mechanical connection between the actin cytoskeleton and
extracellular matrix proteins is provided by transmembrane integrins
appearing in four different states \cite{Lauffenburger99,kuuselaalt}, see
\begin{figure}
    \centering
    \includegraphics[width=0.4\textwidth]{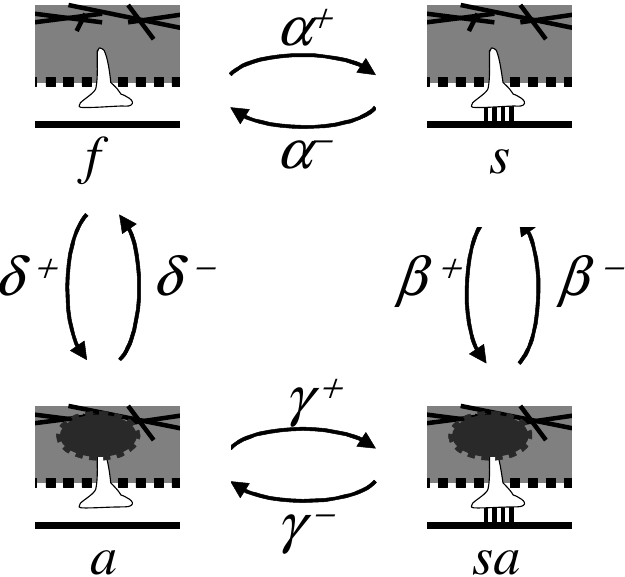} \\
    \caption{Scheme of four states of the transmembrane
      adhesion protein integrin.
      $(f)$: unbound and freely diffusing within the dorsal membrane,
      $(s)$: bound to the substrate,
      $(a)$: bound to the actin network and thus moving with the actin-flow,
      $(sa)$: bound to actin and substrate (force-transducing state).
      Integrins can switch between these states due to reversible binding
      kinetics with binding/unbinding rates
      $\alpha$, $\beta$, $\gamma$ and $\delta$.
    }\label{fig:integrins}
\end{figure}
Fig.~\ref{fig:integrins}.

In the freely diffusing state $f$, integrins are neither coupled to
the substrate nor to the actin cytoskeleton and move according to a
simple diffusion-transport law within the dorsal membrane. These
integrins can change their state by binding to the actin
cytoskeleton $(a)$ or to the substrate $(s)$. In state $a$, the
integrins move with the cortical F-actin velocity $\mathbf v_c$,
whereas integrins in state $s$ remain stationary with respect to the
substrate. Actin- or substrate-bound integrins can switch back into
the freely diffusing state $f$ by unbinding, or into the state $sa$
by coupling to the cytoskeleton and the extracellular matrix. Cell
adhesion occurs only in this double bound state $s$ representing a
focal adhesion (FA), where the frictional force $\mathbf F_c$
(\ref{eq:CorticalFriction}) of the moving
actin-network is transduced to the substrate. \\
The concentration of integrins $c_\#$ in the different states is
described by the following coupled system of differential equations
consisting of terms for spatial movement and binding kinetics.
\begin{align}
\partial_t c_f  &=   \nabla \cdot \bigl( D_f \nabla c_f -  c_f \mathbf u \bigr)
                   + \alpha^- c_s
                   + \delta^- c_a
                   - \bigl( \delta^+_0 a + \alpha^+ \bigr) c_f
                \label{eq:dtcf} \\
\partial_t c_a  &= - \nabla \cdot \bigl( c_a \mathbf v_c \bigr)
                   + \delta^+_0 a c_f
                   - \bigl( \delta^- + \gamma^+ \bigr) c_a
                   + \gamma^-_0 \exp \bigl( 
                       \rho_\gamma |\mathbf F_c| 
                     \bigr) c_{sa}
                \label{eq:dtca} \\
\partial_t c_s  &=   \alpha^+ c_f
                   - \bigl( \alpha^- + \beta^+_0 a \bigr) c_s
                   + \beta^-_0 \exp \bigl( 
                       \rho_\beta |\mathbf F_c| 
                     \bigr) c_{sa}
                \label{eq:dtcs} \\
\partial_t c_{sa} &=   \gamma^+ c_a
                   + \beta^+_0 a c_s
                   - \Bigl( 
                       \beta^-_0  \exp \bigl( \rho_\beta  |\mathbf F_c| \bigr) +
                       \gamma^-_0 \exp \bigl( \rho_\gamma |\mathbf F_c| \bigr)
                     \Bigr) c_{sa}
                \label{eq:dtcsa}
\end{align}
The bonds of the force-transducing integrins ($c_{sa}$) can break
when they experience a mechanical stress by the cytoskeleton, in our
model given by the modulus of the force vector $\mathbf F_c$ defined
in (\ref{eq:CorticalFriction}). Then, according to the theory of
Bell \cite{bell:abriss,Sei00} the dissociation rates $\gamma^-$ and
$\beta^-$ depend exponentially on the mechanical load $|\mathbf
F_c|$, see equations (\ref{eq:dtca}--\ref{eq:dtcsa}) above, with
$\gamma^-_0$ and $\beta^-_0$ describing the basic dissociation rates
without load and the exponential coefficients $\rho_\# =
\zeta_\#/k_B T$ measuring potentially different rupture rates from
the substrate or cytoskeleton binding site, respectively.

\subsubsection{Mass flux conditions at the free boundary.}
In addition to the already mentioned bulk flux conditions
(\ref{eq:NormalFlow}) and (\ref{eq:CorticalFlow}) related to the
normal speed of the cell edge $\Gamma(t)$, we have to impose
compatible boundary conditions onto the concentrations of those
molecular species that are not fixed to the substratum. For the two
parabolic diffusion equations (\ref{eq:dtmf}) and (\ref{eq:dtcf}) we
impose, respectively, natural zero-flux boundary conditions onto
freely diffusing myosin ($m_f$) and fixed Dirichlet conditions ($c_f
= c_f^0$) onto freely diffusing integrin. Thereby we suppose a
constant reservoir of fresh adhesion receptors expressed in the
upper ventral membrane and diffusing (or eventually being
transported) from there around the lamellar tip into the lower
dorsal part of the membrane.

\begin{figure}[htbp]
  \centering
  \includegraphics[width=0.5\textwidth]{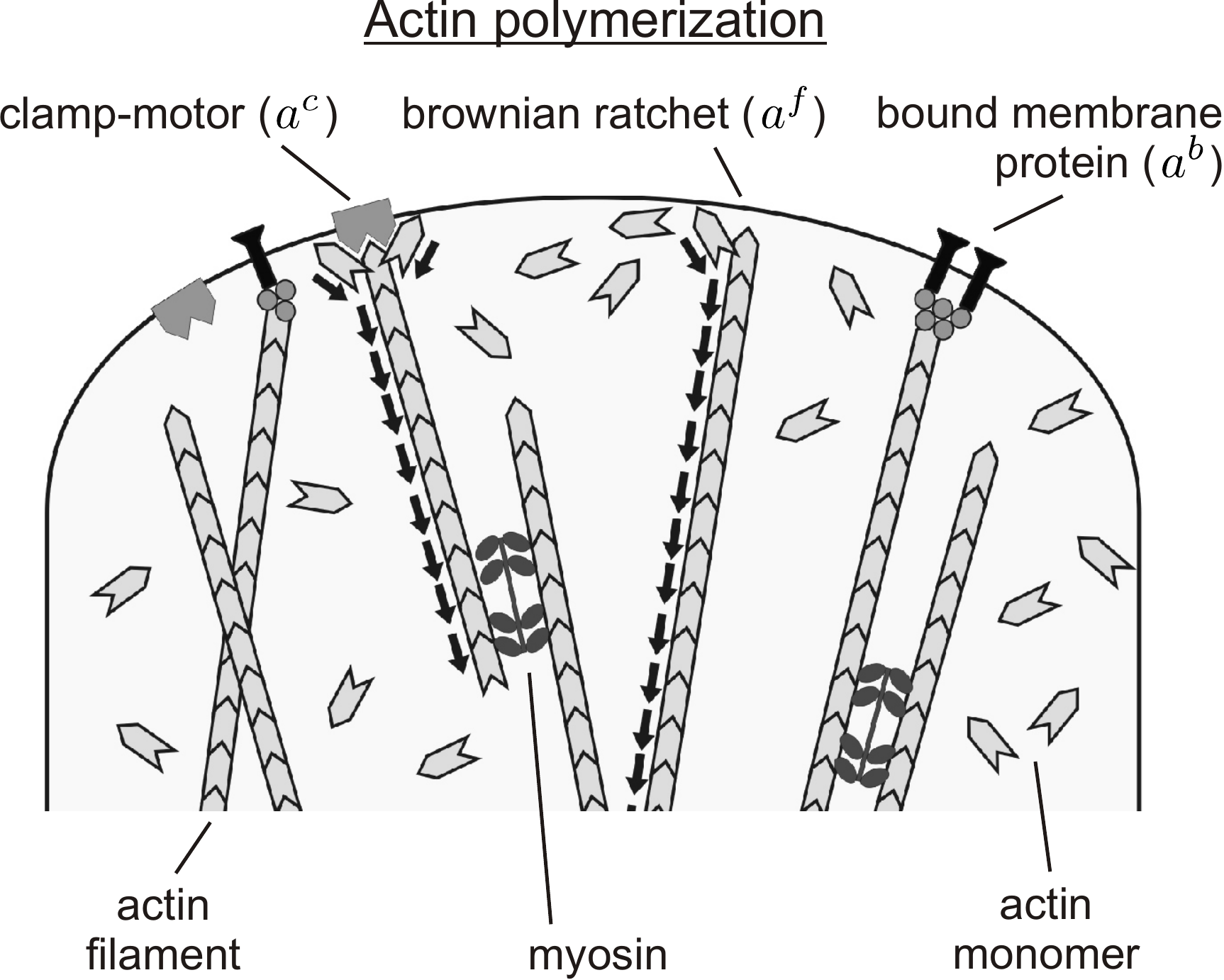} \\
    \caption{Schematic top view onto the lamellar tip of a cell
             (fragment) showing three possibilities for
             actin filament ends to interact with the free plasma
             membrane: anchoring at a membrane protein, as integrin;
             freely fluctuating with polymerization by a 
             ``Brownian ratchet'' mechanism;
             binding to a ``clamp-motor'' protein, like the WASP-complex, 
             with induced polymerization.
             The resulting local pseudo-equilibrium F-actin concentrations 
             are $a^b, a^f$, and $a^c$, respectively.
             See equations (\ref{eq:Ratchet}) and
             (\ref{eq:Polymerization}) for a quantitative model.
    }\label{fig:tip} %
\end{figure}

The F-actin flux $\theta \mathbf v$ cannot leave the cell, meaning
that on the free boundary $\Gamma$ the \emph{relative inward normal
F-actin velocity} always has to satisfy the inequality
\begin{equation}
\label{eq:InwardVelocity} V = \dot{\Gamma} - \nu_\Gamma\cdot\mathbf v \geq 0.
\end{equation}
If the strong inequality $V>0$ holds at a certain boundary point,
then two different modeling situations may arise for the cell edge:
\begin{enumerate}
\item {\emph{No-stick condition at the lamellar tip:}}
  Local disruption of the actin network from the edge is allowed,
  e.g.~under suitable load conditions \cite{kuuselaalt}, so there
  is no new F-actin production directly at plasma membrane edge.
  Therefore we have to impose zero-Dirichlet condition for the F-actin
  concentration
  \begin{equation}
  \label{eq:ThetaZero} \theta = 0 \quad \text{if} \; V > 0 \; 
  \text{holds at a non-sticky point} \; \mathbf x \in \Gamma.
  \end{equation}
\item {\emph{Network sticking to the lamellar tip:}}
  As indicated in Fig.~\ref{fig:tip}, active polymerization of actin filaments
  directly at the plasma
  membrane is allowed either (a) at fluctuating free filament ends touching the
  tip membrane \cite{mogilner:motility},
  or (b) at filaments that are bound to
  clamp-motor proteins anchored in the tip membrane
  \cite{dickinson:polymerMotors}.
  Both cases could occur simultaneously in a
  local region, but whether active polymerization with inward mass
  flux $\theta V>0$ can take place depends on local force balance
  conditions, see Section \ref{sec:forcebalanceeqs}
  and (\ref{eq:PolyNeumann_v}) below.
\end{enumerate}

Finally, also the two hyperbolic equations (\ref{eq:dtmb}) and
(\ref{eq:dtca}) require zero-influx conditions, so that for the
transported concentrations $m_b$ and $c_a$ we have to impose
zero-Dirichlet conditions only in cases of $V>0$ or $V_c>0$,
respectively. In the latter case the inward normal velocity of
$\mathbf v_c$ is defined in analogy to (\ref{eq:InwardVelocity}),
noting that under the modeling hypothesis (\ref{eq:MembraneFlow}) we
anyway have $V_c=\kappa V$.

\subsection{Force balance equations}\label{sec:forcebalanceeqs}

\subsubsection{Two-phase flow equations for cytoskeleton and cytosol.}
Since cell movement usually occurs on a time scale of minutes, and
cell sizes usually are in the order of tens of microns, already the
cytosol, with its consistency similar to an aqueous viscous fluid,
has low Reynolds number. The more this is true for the cytoskeleton
with consistency similar to a dense viscoelastic gel. The
viscoelastic properties of the cytoplasm have been experimentally
studied \cite{evans:viscosity,Feneberg01} and also mathematically
modeled \cite{dembo1986,Oli05}, whereby the simple model of a
Maxwell fluid seems to serve as a quite good description. This
means, that elastic properties are mainly effective on a shorter
time scale of seconds and dominated by viscous effects on a medium
scale of minutes. Moreover, in contrast to a passive elastic
material, the cytoplasm mostly stays under active contractile stress
as exerted by myosin-II oligomers (in the order of kPa), which is
able to break weaker cross-links of the F-actin network e.g.~by
$\alpha$-actinin or filamin, so that the contraction forces are
equilibrated only by viscous and frictional forces.

Under these assumptions, the highly viscous two-phase \emph{creeping
flow model} for cytoplasm, originally proposed 25 years ago by Dembo
et al.~\cite{DemboAl1984} and so far extensively applied, see
e.g.~\cite{wolfgang:biomech,Her03}, can be condensed into a
pseudo-stationary linear elliptic system for the \emph{mean F-actin
velocity} $\mathbf v$ and the \emph{effective hydrostatic pressure}
$p$ on $\Omega(t)$:
\begin{align}
\label{eq:Stokes}
  \nabla \cdot \mu \theta \widetilde{\nabla} \mathbf v +
  \nabla \Bigl( S(\theta,m_b) - p \Bigr) &=
      \Phi(\theta,c_{sa}) \mathbf v_c \, ,\\
\label{eq:Darcy} \nabla p &= \varphi \theta ( \mathbf v - \mathbf
w)\, .
\end{align}
Here the generalized (elliptic) Stokes equation (\ref{eq:Stokes})
involves the effective stress $S(\theta,m_b)$ as defined in equation
(\ref{eq:Stress}) and the symmetrized displacement rate $\widetilde
\nabla \mathbf v$ (see e.g.~\cite{kuuselaalt}). Moreover, $\mathbf
v_c$ is defined in (\ref{eq:CorticalFlow}), and $\mu, \Phi$ denote
the coefficients of bulk viscosity and substratum friction, see
(\ref{eq:Phthcsa}). Because of negligible cytosol viscosity the
simple Darcy law (\ref{eq:Darcy}) suffices as a model description,
where the coefficient $\varphi$ measures the internal two-phase flow
friction. From the last equation one explicitly solves for the
cytosol velocity $\mathbf w = \mathbf v - (1/\varphi\theta) \nabla
p$, so that the total volume flux is
\begin{equation}
\label{eq:Welocity} \mathbf W = \mathbf v + (1-\theta)(\mathbf w-\mathbf v) =
                          \mathbf v - \frac{1-\theta}{\varphi\theta} \nabla p.
\end{equation}
Insertion into the mass-balance equation (\ref{eq:DivergenceFree})
then yields the generalized (elliptic) Laplace equation
\begin{equation}
\label{eq:Laplace} \nabla\cdot\frac{1-\theta}{\varphi\theta}\nabla p =
    \nabla\cdot \mathbf v.
\end{equation}
Notice that here the ellipticity degenerates for marginal volume
fractions $0 < \theta < 1$, which can be relevant in cases where
$\theta\rightarrow0$ at boundary points of cytoskeleton disruption,
see (\ref{eq:ThetaZero}) above.

We remark that together with the hyperbolic mass balance equation
(\ref{eq:MassBalance}) the linear elliptic system
(\ref{eq:Stokes},\ref{eq:Laplace}) constitutes generalized
pseudo-stationary Navier-Stokes equations for the F-actin network as
a compressible, highly viscous and reactive fluid. It is mutually
coupled to the mass concentrations in equations (\ref{eq:dtmb}) and
(\ref{eq:dtcsa}) via a myosin-mediated contractile stress term
appearing in $S(\theta,m_b)$, see eq.~(\ref{eq:Stress}) below, and
an adhesion-mediated friction coefficient
\begin{equation}\label{eq:Phthcsa}
\Phi(\theta,c_{sa}) = \Phi_0 c_{sa} \theta.
\end{equation}
In this way the right hand side of (\ref{eq:Stokes}) reads $\mathbf
F_v = \Phi \mathbf v_c = c_{sa} \mathbf F_c$, with the frictional
force $\mathbf F_c$ per doubly bound integrin defined in
(\ref{eq:CorticalFriction}) above.

Tracing these model equations back to their derivation
\cite{dembo1989,Alt03,kuuselaalt} provides us not only with precise
biophysical conditions on $\mathbf v$ and $p$ at the moving boundary
$\Gamma$ (see next paragraph), but also with genuine nonlinear
parameter functions being based on thermodynamical reasoning at the
molecular scale: The function $S$ in (\ref{eq:Stokes}) represents
the effective stress in the network phase, which is induced by
molecular interactions between the F-actin filaments themselves as
well as by thermic interactions with solvent molecules in the
cytosol. It is generally expressed as the weighted negative sum $S =
- \theta P_a - (1-\theta)P_s$ of corresponding pressures $P_a$ and
$P_s$ which are applied to any network volume element and any
cytosol element, respectively.
Passive elastic stresses may be neglected since these are already
relaxed on the creeping flow timescale. When finding a free binding
site on filaments, previously single bound myosin-II tetramers exert
power stroke motor forces with their free myosin heads,
cf.~Figs.~\ref{fig:lamellipodium} and \ref{fig:tip}. Thus, an
attractive stress $- P_a = \psi_0 m_b$ is applied, where the simple
coefficient $\psi_0$ comprises binding affinity, power stroke
probability and the mean applied force per stroke. On the other
hand, standard Gibbs free energy arguments suggest a molecular
\emph{solvent pressure} $P_s = - \sigma_0 \ln(1-\theta)/(1-\theta)$,
see \cite{Alt03}. Then we arrive at expressions for the
\begin{align}
\label{eq:Stress}
  \text{\emph{effective stress}}& &
  S = S(\theta,m_b) &= \psi(\theta,m_b) - \sigma(\theta)
  &\phantom{-}& \\
\label{eq:Contraction}
  \text{\emph{contractile stress}}& &
  \psi(\theta,m_b) &= \psi_0 \theta m_b
  &\phantom{-}& \\
\label{eq:Swelling}
  \text{\emph{swelling pressure}}& &
  \sigma(\theta) &= \sigma_0 |\ln(1-\theta)|,
  &\phantom{-}&
\end{align}
see
\begin{figure}
  \begin{minipage}[b]{0.5\textwidth}
    \includegraphics[width=1.0\textwidth]{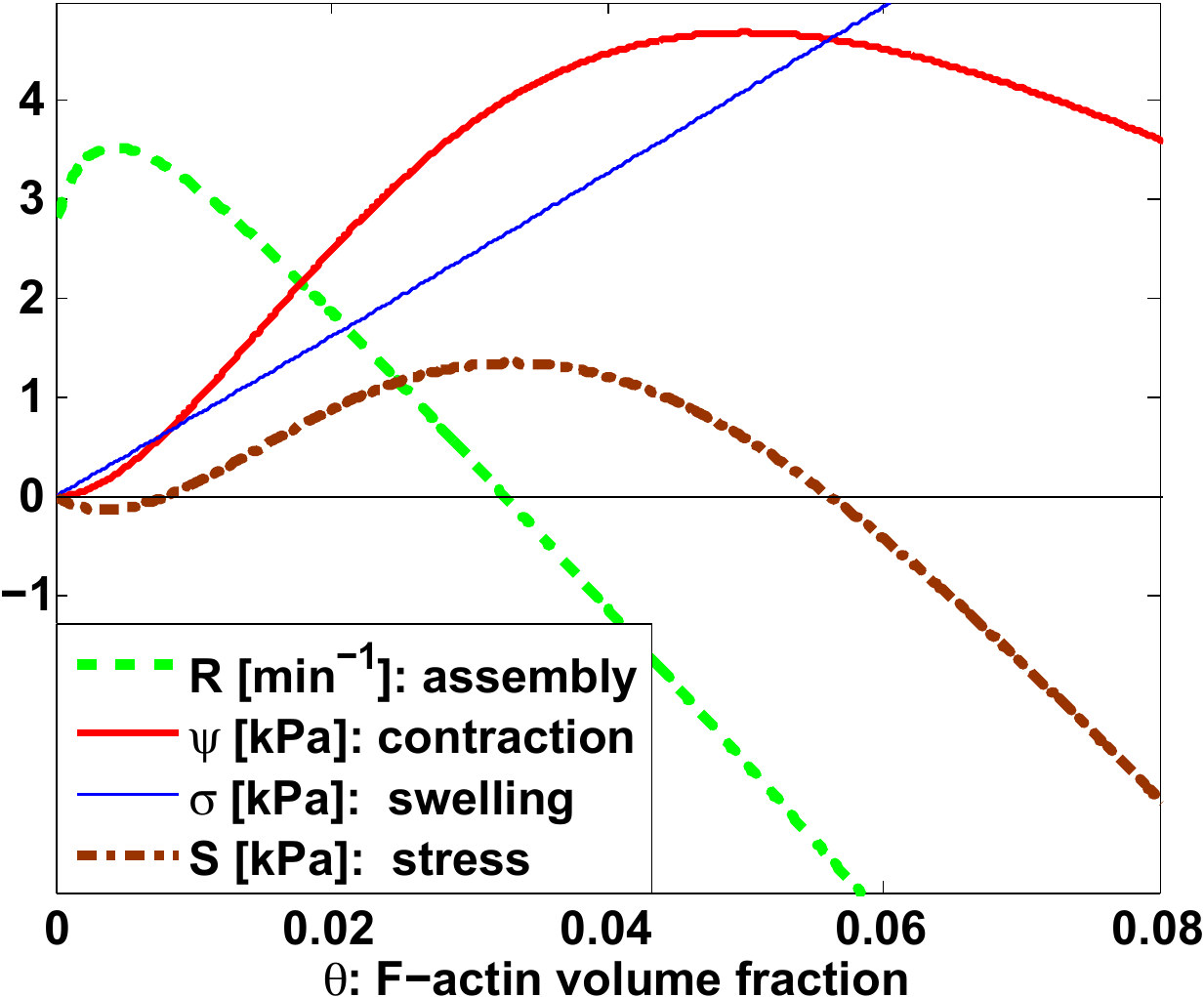}
  \end{minipage} \hskip 2ex
  \begin{minipage}[b]{0.47\textwidth}
    \caption{Plots of F-actin model functions.
             Dashed curves: net assembly rate
             $R (\theta)$, equation (\ref{eq:Assembly}), and effective stress
             $S(\theta)=\psi(\theta) - \sigma(\theta)$, equation
             (\ref{eq:Stress}).
             Continuous curves: contractile stress $\psi$ defined in
             equation (\ref{eq:ContractionTheta})
             and swelling pressure $\sigma$, equation (\ref{eq:Swelling}).
            }\label{fig:assembly}
  \end{minipage}
\end{figure}
Fig.~\ref{fig:assembly} for corresponding plots with $m_b=m_b^*$
in equation (\ref{eq:MyosinStar}). We emphasize that the effective stress
function in equation (\ref{eq:Stress}) sums up the contributions from the
cytoskeleton and the cytosol, so that both phases are not separated
anymore. However, when deriving the corresponding natural boundary
conditions for the elliptic problem, then the two phases have to be
considered separately again.

\subsubsection{Stress and pressure balance conditions at the free boundary.}
Recall that according to (\ref{eq:NormalFlow}) the cell
edge $\Gamma(t)$ always moves together with the normal component of
the total cytoplasm flux $\mathbf W$ as quantified in (\ref{eq:Welocity}).
Then, from (\ref{eq:InwardVelocity}) also the
relative inward normal F-actin flux is determined as proportional to
the inward normal pressure gradient on $\Gamma(t)$
\begin{equation}
\label{eq:NormalGradient}
  \theta V = \theta \Bigl( \dot{\Gamma} - \nu_\Gamma\cdot\mathbf v \Bigr) =
            -\frac{1-\theta}{\varphi}\nu_\Gamma\cdot\nabla p.
\end{equation}
This explicit linear relation between free boundary speed, normal
F-actin velocity and pressure gradient serves as an extra free
boundary condition in addition to the set of boundary conditions
that are necessary for uniquely solving the linear elliptic system
(\ref{eq:Stokes}) and (\ref{eq:Laplace}) on the given domain
$\Omega(t)$. These depend on the particular model choice of boundary
pressure functions at the cell edge membrane, but also on the
outcome of the F-actin flow requirement $V\geq0$ in
(\ref{eq:InwardVelocity}).

Using the general derivations in \cite{Alt03}, eqs.~(58-62), one
obtains separate \emph{pressure balance} conditions for each of the two
phases at all free boundary points of $\Gamma(t)$ satisfying $V>0$,
\begin{align}
\text{\emph{cytoskeleton}}& \label{eq:BdryStress_a}&
  - \nu_\Gamma \cdot \mathbb T_a \cdot \nu_\Gamma + \theta p =&
                                                      \theta P_a^\Gamma \\
\text{and \emph{cytosol}}& \label{eq:Dirichlet_p}&
  (1-\theta) p + \sigma(\theta) =& (1-\theta) P_s^\Gamma&
\end{align}
with the intrinsic stress tensor
$ \mathbb T_a = 
  \mu \theta \widetilde{\nabla} \mathbf v + \psi (\theta,m_b) \mathbb I$.
These equations mean that at those parts of the tip plasma membrane,
which are exposed to the cytoskeleton network ($\theta$) or to the
cytosol ($1-\theta$), respectively, 
the sum of internal pressures
is in balance with a certain \emph{boundary cytoskeleton pressure}
$P_a^\Gamma$ or \emph{boundary cytosol pressure} $P_s^\Gamma$ at
each volume element. Modeling expressions for these pressures will
be given in the following paragraph.

Summing up both pressure balance equations (\ref{eq:BdryStress_a})
and (\ref{eq:Dirichlet_p}) yields the general \emph{Neumann-type
condition} for $\mathbf v$ on $\Gamma(t)$
\begin{equation}
\label{eq:Neumann_v}
  \nu_\Gamma \cdot \mathbb T_a \cdot \nu_\Gamma -
  \bigl( \sigma(\theta) + p \bigr) +
  \theta P_a^\Gamma +
  (1-\theta) P_s^\Gamma = 0,
\end{equation}
which is now valid without restriction, see \cite{Alt03}, eq.~(58),
i.e.~also in boundary points where the condition $V=0$ holds.
However, then the Dirichlet-type boundary condition
(\ref{eq:Dirichlet_p}) for $p$ has to be replaced by a corresponding
inequality, and insertion of $V=0$ into (\ref{eq:NormalGradient})
provides a \emph{Neumann condition} for $p$ instead. In particular
we have $\nu_\Gamma\cdot\nabla p = 0$, which also holds in case of
disruption with $V>0$ and $\theta=0$.

Thus, in any constellation of the mentioned conditions on
$\Gamma(t)$, for fixed time $t$ and given profile of all protein
concentrations, we obtain a well-posed elliptic boundary value
problem for $\mathbf v$ and $p$ to be solved on $\Omega(t)$ (eventually by
numeric iteration and/or conjugate gradient method,
cf.~\cite{kuuselaalt}). Finally, the normal speed $\dot{\Gamma}$ of
the free boundary can be calculated by using equation
(\ref{eq:NormalGradient}). This is also possible in the case of
network disruption at some boundary point $\mathbf x_0 \in \Gamma(t)$ with
$\theta(t,\mathbf x_0)=0$, since then $V(t,\mathbf x_0)>0$ is obtained as the
L'Hospital-limit of
$-\frac{1-\theta(t,\mathbf x)}{\theta(t,\mathbf x)}\nu_\Gamma\cdot\nabla
p(t,\mathbf x)$ as $\mathbf x \rightarrow \mathbf x_0$ from the interior.

\subsubsection{Boundary pressure functions at the cell edge.}
We only consider creeping cell migration in a medium that is not
under external pressure or stress. Therefore, in a purely physical
model, the boundary pressures $P_a^\Gamma$ and $P_s^\Gamma$ along
the cell edge $\Gamma(t)$ should be set to zero. However, the edge
is physically defined as the lamellar tip, where the ventral and
dorsal part of the surrounding plasma membrane meet. Thus,  if
$\tau_\Gamma$ denotes the scalar tension value of the dorsal plasma
membrane at the cell edge (see the following paragraph) then this
tension appears as a \emph{boundary tip pressure} acting on both
phases, the cytoskeleton and the cytosol. Moreover, in addition to
the counteracting cytoskeleton and cytosol pressure at the left hand
sides of eqs.~(\ref{eq:BdryStress_a}) and (\ref{eq:Dirichlet_p}),
there could be  extra pressures due to \emph{active polymerization
forces}, which usually are generated at barbed actin filament ends
that can become exposed to the tip membrane in two variants, see
Fig.~\ref{fig:tip} above:
\begin{itemize}
\item {\bf Brownian ratchet model:}
  Assume that at a point of the lamellar tip $\Gamma(t)$
  a fraction $a^B$ of filaments is bound to membrane proteins
  in a fast pseudo-steady state equilibrium
  with the actual F-actin concentration $a = \theta a_\mathrm{max}$.
  Then from the remaining network with concentration $a^F(a) = a
  - a^B(a)$, Arp2/3 induced branching can occur. The barbed
  filament ends are more or less normally exposed to the tip
  membrane with concentration
  $a^f = a^f (a) = \alpha_0^f \cdot \mathrm{Arp0} \cdot \frac{a^F}{K_a + a^F}$
  with suitably chosen coefficients, see also (\ref{eq:BarbedEnds}).
  Then, due to insertion of G-actin
  in-between fluctuating filament and membrane there appears a
  \emph{free polymerization pressure}
  \begin{equation}
  \label{eq:Ratchet} p^f = \pi_0^f \cdot a^f(a)
  \end{equation}
  with a \emph{ratchet coefficient} $\pi_0^f>0$ depending on the
  free energy of one monomer addition, see relevant force
  estimations e.g.~in \cite{mogilner:cellMotility}.
\item {\bf Clamp-motor model:}
  Alternatively or additionally, a certain fraction of tip bound actin
  filaments with concentration $a^c = a^c(a) < a^B(a)$ can be bound to
  WASP-like membrane proteins,
  which serve as filament-end-tracking motors.
  By means of an energy consuming polymer elongation process
  at the clamped end of the filament, these proteins
  push the bound filament outward,
  thus leading to a \emph{clamp polymerization pressure}
  \begin{equation}
  \label{eq:Ratchet2} p^c = \pi_0^c \cdot a^c(a).
  \end{equation}
  Here again, the clamp-motor coefficient $\pi_0^c>0$ depends on
  the energy of one monomer translocation, see
  \cite{dickinson:polymerMotors}.
\end{itemize}
The sum of these active polymerization pressures would induce an
averaged relative normal inward mass flux $\theta V \geq 0$ of the
whole cytoskeleton, based on the particular inflows of $a^f$- and
$a^c$-filaments satisfying $\theta V=a^f V_f = a^c V_c$. However,
this flow experiences a viscous resistance from the remaining fixed
filaments (concentration $a^B - a^c$), with a viscosity coefficient
that could increase in the presence of bound myosin molecules. Thus
we get the total \emph{tip polymerization pressure}
\begin{equation}
\label{eq:Polymerization}
P_\mathrm{poly} = p^f + p^c - \mu_\Gamma(m_b) \bigl( a^B - a^c \bigr) a_\mathrm{max} V.
\end{equation}
Clearly, here the relative inward velocity $V$ is the one defined by
the normal component of $\mathbf v$ in relation to $\dot\Gamma$, see
(\ref{eq:InwardVelocity}). For a similar resistance model based on
elastic cross-linking see \cite{gholami:oscillations08}.

Finally, since this intrinsic boundary pressure $P_\mathrm{poly}$ acts onto
the cytoskeleton volume fraction, while simultaneously inducing a
corresponding counter-pressure onto the cytosol volume fraction, we
can state the following distribution for the boundary pressures
generated at the membrane
\begin{align}
\label{eq:ExtPressure_a}
  \theta P_a^\Gamma &= \theta (1-\kappa_\Gamma) \tau_\Gamma + P_\mathrm{poly}\, , \\
\label{eq:ExtPressure_s}
  (1-\theta)  P_s^\Gamma &= (1-\theta +\theta\kappa_\Gamma) \tau_\Gamma -
  P_\mathrm{poly}\, ,
\end{align}
where we introduce a weight factor, $0\leq\kappa_\Gamma<1$,
measuring the relative effect of membrane tension $\tau_\Gamma$ onto
the cytosol phase and possibly depending on the not-explicitly
modeled tip geometry. Then, by substituting $p$ from
(\ref{eq:Dirichlet_p}) into (\ref{eq:Neumann_v}) we obtain the
generalized \emph{Neumann-type boundary condition} for $\mathbf v$
in all points of $\Gamma(t)$ where the network is attached, $a^B >
0$, and where $V>0$ holds:
\begin{equation}
\label{eq:PolyNeumann_v}
  \nu_\Gamma \cdot \mathbb T_a \cdot \nu_\Gamma - \sigma(\theta) -
  \frac{a_\mathrm{max}}{1-\theta} \mu_\Gamma(m_b) \Bigl( a^B - a^c \Bigl) V +
  \frac{1}{1-\theta} \Bigl(
    \sigma(\theta) + p^f + p^c -\theta \kappa_\Gamma \tau_\Gamma
  \Bigr)
  = 0.
\end{equation}
This means, that the \emph{mean inward F-actin polymerization speed}
$V\geq 0$ on the moving cell edge $\Gamma(t)$ is implicitly
determined by solving the linear elliptic system (\ref{eq:Stokes})
and (\ref{eq:Laplace}) for $\mathbf v$ and $p$ and satisfying all
boundary conditions. Thereby the Neumann condition above contains
all membrane protruding pressure terms in series, namely the
swelling pressure $\sigma$, the polymerization pressures $p^f$ and
$p^c$ induced by a Brownian ratchet or an clamp-motor mechanism, as
well as a counteracting stress due to dorsal membrane tension
$\tau_\Gamma$.

\subsubsection{Global force balance at the adhesive substratum.}
There are two kinds of forces exerted by the migrating cell
(fragment) onto the fixed flat substratum: the integrin-adhesion
mediated active frictional force represented by the vector field
$\mathbf F_v$ on the right hand side of Stokes' equation
(\ref{eq:Stokes}), and an analogous passive frictional force $\mathbf F_u$
due to motion of the dorsal plasma membrane along the substratum:
\begin{align}
  \label{eq:AdhesionForce}
  \mathbf F_v &= c_{sa} \mathbf F_c = \Phi ( \theta, c_{sa} ) \mathbf v_c, \\
  \label{eq:FrictionForce}
  \mathbf F_u &= \Phi_u \mathbf u,
\end{align}
with $\Phi$ as in (\ref{eq:Phthcsa}) and an assumed friction
constant $\Phi_u \geq 0$. Supposing that on the substratum no other
forces are applied than these, then their integral sum has to
vanish, so that the zero force balance holds:
\begin{equation}
\label{eq:ZeroForce}
0 = \int_{\Omega(t)}(\mathbf F_v + \mathbf F_u).
\end{equation}
Furthermore, if we assume that the dorsal membrane, viewed as a
2-dimensional incompressible fluid satisfying the zero-divergence
condition (\ref{eq:MembraneFlow}), has relatively low
viscosity, then the \emph{membrane tension} $\tau_u$ induced by the
frictional flow can be defined according to Darcy's law
\begin{equation}
\label{eq:Tension_u} \nabla \tau_u = \Phi_u \mathbf u
\end{equation}
and thus determined as solution of the Laplace equation $\Delta
\tau_u = 0$ with Neumann boundary condition $\nu_\Gamma\cdot\nabla
\tau_u = 0$, see (\ref{eq:MembraneFlow}). Then the zero force
balance (\ref{eq:ZeroForce}) together with (\ref{eq:Stokes})
implies that the boundary tension values $\tau_\Gamma =
\tau_u|_{\Gamma(t)}$, which are uniquely determined up to a
constant, necessarily fulfill the \emph{integrability condition}
$\int_{\Gamma(t)} ( \mu\theta \widetilde{\nabla} \mathbf v +
                    S(\theta,\mu_b) - p + \tau_\Gamma )
                  \nu_\Gamma = 0$,
which by insertion of the Neumann boundary condition
(\ref{eq:Neumann_v}) and by using the symmetry of
$\widetilde{\nabla} \mathbf v$ 
reduces to the equivalent necessary condition
$\int_{\Gamma(t)} ( \theta P_a^\Gamma + (1-\theta) P_s^\Gamma -
                    \tau_\Gamma )
                  \nu_\Gamma = 0$.
Indeed, this condition is fulfilled for the boundary pressure model
functions that were chosen in the preceding paragraph,
(\ref{eq:ExtPressure_a}) and (\ref{eq:ExtPressure_s}), because then
even the integrand in the previous condition vanishes.

\section{Results of model simulations}

\subsection{Spontaneous cell polarization in the 2-D model}\label{sec:spontan}

We have simulated the adhesive motion of a flat cell or cell
fragment represented by a 2-dimensional domain, $\Omega(t)$, with
moving cell edge or lamellar tip, $\Gamma(t) = \partial\Omega(t)$,
under certain simplifying assumptions:
\begin{enumerate}
  \item The lower dorsal membrane sticks to the substratum ($\mathbf u=0$) and
        there exists a small membrane tension $\tau_\Gamma > 0$,
        constant over the whole cell edge.
  \item There occurs no active polymerization pressure at the cell edge
        ($p^f = p^c = 0$), only the similar swelling pressure
        $\sigma(\theta)$ and the hydrostatic pressure $p$
        can push the boundary.
  \item Disruption of the F-actin network from the lamellar tip
        $\Gamma(t)$ can locally occur if the network tension exceeds a
        certain threshold that might depend on the fraction of membrane-bound actin
        filaments $a^B = A \frac{a}{K_B + a}$.
  \item The free myosin-II concentration is at a fixed constant level
        $m_f^0>0$ and the amount of F-actin bound myosin-II oligomers is in
        a pseudo-steady equilibrium $m_b^*(a)$ according to
        (\ref{eq:MyosinStar}), so that the contractile stress is only a
        function of $\theta = a/a_\mathrm{max}$:
        \begin{equation} \label{eq:ContractionTheta}
        \psi(\theta)=\psi_0 \theta m_b^* =
           \frac{\psi_0\alpha_m m_f^0}{a_\mathrm{max}\delta_m^0}
                            \frac{\theta^2}{1 + \theta^2/\theta_\mathrm{opt}^2}.
        \end{equation}
\end{enumerate}

More details and a list of chosen parameters can be found in
\cite{kuuselaalt} (Section 1 and Table 2). Since unpolarized cells
and cell fragments, as observed under various conditions
\cite{verkhovsky:polarization,pmid17893245}, attain a quite regular
circular shape, we choose as initial condition a circle $\Omega(0)$
of radius $6 \, \mu \mathrm m$. Moreover, in order to mimic the radial
spreading of cells after exposure onto a flat substratum, we start
with constant integrin densities and radially symmetric initial
configuration for the volume fraction $\theta$, with slightly larger
values closer to the center.  Due to this initial perturbation, the
F-actin concentration rapidly condenses into a central region of
high $\theta$, surrounded by a lamella-kind region of low $\theta$,
see
\begin{figure}[htbp]
  \centering
  \includegraphics[width=0.8\textwidth]{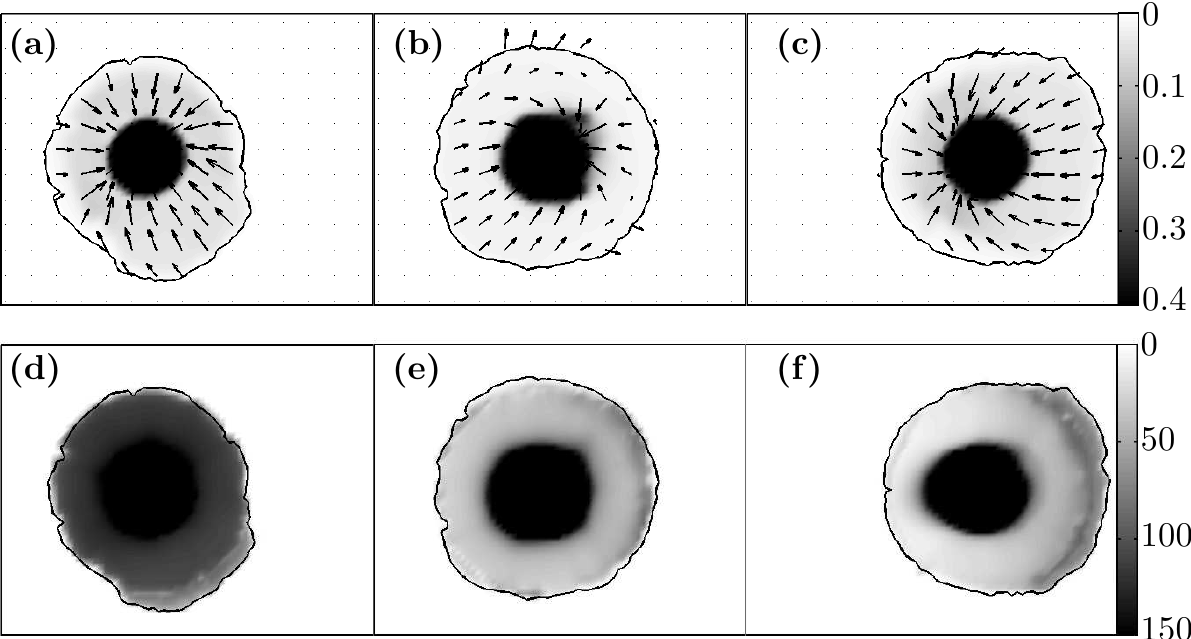}
  \caption{2-D simulation of a spontaneously polarizing
           cell (fragment) starting
           with radially symmetric initial conditions:
           Spatial distributions shown as pseudocolor plots of
           F-actin volume fraction $\theta$ in (a,b,c),
           and concentration of actin-and-surface bound integrin
           proteins $c_{sa}$  in (d,e,f)
           at three different time instants:
           (a,d) 1 min, (b,e) 5 min, and (c,f) 10 min after
           initialization.
           The width of the shown region is $22 \, \mu \mathrm m$.
          }\label{fig:cellFragment}
\end{figure}
Fig.~\ref{fig:cellFragment}(a,b). Later, also the actin- and
surface-bound integrin adhesion proteins $c_{sa}$ concentrate around
this center, see Fig.~\ref{fig:cellFragment}(e). Both phenomena are
supported by a strong retrograde F-actin flow, which collects actin
filaments and actin bound integrins $c_a$ in radial direction from
almost everywhere in the periphery. Moreover, the hydrodynamic
pressure has its maximum within the center region (data not shown),
so that its negative outward gradient represents the squeezed flow
of cytosol out of the contracting F-actin network.

At the free cell edge with relatively low F-actin concentration,
there occurs a meta-stable equilibration between a positive swelling
pressure pushing the lamellar tip outwards and a resisting viscous
network tension pulling the tip inwards. For some time, see
Fig.~\ref{fig:cellFragment}(b) after $5$ minutes, local regions of
protrusion or retraction can be observed, which point into varying
directions along the cell periphery. Notice that these
spatiotemporal fluctuations are not due to the tiny stochastic
perturbations that are imposed to the F-actin polymerization rate, 
but represent the emergent chaotic dynamics of the cytoplasm as a
reactive and contractile two-phase fluid \cite{dembo1986,altdembo}.
Furthermore, behind a cyclically protruding and retracting free edge
we can observe a layer with slightly increased concentration of
substrate-and-actin bound integrin $c_{sa}$, see
Fig.~\ref{fig:cellFragment}(e). This is the onset of polarization:
fresh free integrin proteins are appearing at the protruding part of
the edge from the upper membrane, thus also increasing $c_{sa}$ in
this region. As a consequence, a larger frictional force of the
retrograde F-actin is generated in that direction, inducing a bias
into the force vector field transduced to the substratum. Thus, the
whole cell fragment starts to move in this direction, which later
becomes the leading edge of the migrating cell fragment, see
Fig.~\ref{fig:cellFragment}(c,f). The emerging polarization of the
cell is most clearly expressed in the $c_{sa}$ distribution of
Fig.~\ref{fig:cellFragment}(f), where an increasingly dense band
is formed behind the leading edge and the central region of focal
adhesions is slowly shifted rearwards with increasing migration
speed, while becoming deformed in shape similarly to the whole cell
fragment.

This simulation is an example for the autonomous formation of a
meta-stable unpolarized, almost circular state and its spontaneous
transition into a polarized, migrating state of a flat cell
fragment. By changing some of the model parameters we can influence
the degree of this symmetry-breaking instability, but so far we were
not able to reproduce the observed longer-time stability of circular
cell fragments, see again \cite{verkhovsky:polarization}. One reason
for this failure seems to be, that in our model simplifications we
assumed a constant distribution of freely diffusing myosin-II
oligomers, contradicting experimental results on clearly expressed
gradients of myosin-II concentration decaying towards the cell edge,
see e.g.~\cite{svitkina:myosin}. Therefore we have started to
investigate the full coupled model system by including active tip
polymerization and the kinetics for myosin-II diffusion, binding and
transport, in addition to the already implemented analogous kinetics
and dynamics of integrin adhesion molecules. The next section
presents the so far achieved results in the most simple but
nevertheless quite instructive 1-dimensional situation.

\subsection{Induced onset of cell polarization and migration in the
            1-D model}\label{sec:induced}

We have simulated adhesion, polarization and migration of an
idealized flat cell fragment having a fixed extension and only one
degree of freedom to move, as can be observed experimentally, for
example in Fig.~\ref{fig:expFragmente}(c). Thus, in a 1-D
cross-section along its moving direction, the fragment is
represented by an interval $\Omega(t) = [x_b(t),x_b(t)+L]$ of fixed
length $L$, moving with body speed $v_b(t) = \dot{x_b}(t)$, so that
the whole kinetics and dynamics within cytoplasm and dorsal membrane
can be described by the corresponding 1-dimensional equations and
conditions as in Section \ref{sec:forcebalanceeqs}, but now written
in cell-centric coordinates, for example $\tilde{\theta}(t,y) =
\theta(t,x_b(t)+y)$ or $\tilde{v}(t,y) = 
v(t,x_b(t)+y) - v_b(t)$. Then, assuming the following particular
restrictions, we have solved a simplified system of boundary value
problems:
\begin{enumerate}
\item The actin network is always sticky at the lamellar tips
      (no disruption), so that $\tilde{\theta}
      > 0$ on the whole closed interval $[0,L]$.
\item We assume active tip polymerization with a simultaneous
      parallel effect of the brownian ratchet 
      ($\pi_0^f = 5 \, \mathrm{Pa} / \mu \mathrm M$)
      and the clamp-motor mechanism 
      ($\pi_0^c = 5 \, \mathrm{Pa} / \mu \mathrm M$). 
      Moreover, we suppose
      that the membrane-bound cortex shear viscosity strongly increases if
      myosin-II oligomers are bound: 
      $\mu_\Gamma(m_b) = 0.1 (1+45 m_b) \, 
          \mathrm{Pa} \ \mathrm{min} / \mu \mathrm m$.
      For the actin-binding membrane proteins at the tip with maximal
      concentration $A=50 \, \mu \mathrm M$ and self-enhanced binding with
      dissociation constant $K=158.1 \, \mu \mathrm M$ 
      we use the pseudo-equilibrium
      $2a^B (a) = A - \sqrt{(A-a)^2 + 2(A+a)K^2/a
      + (K^2/a)^2} + K^2/a + a$ and $a^c=0.1\ a^B$.
\item The dorsal membrane moves together with the cell (no slip
      at the tips), so that
      $\tilde{u} \equiv 0$ and $v_c = v = \tilde{v} + v_b$.
      The tip membrane tension difference is
      $[\tau_\Gamma]_0^L = \Phi_u L v_b$ with the minimum always equal to
      a fixed positive constant $\tau_0=25 \, \mathrm{Pa}$. 
      Finally, the substrate force
      balance (\ref{eq:ZeroForce}) reads
      \begin{equation}
      \label{eq:SubstrateForce} \Phi_u L v_b =
        \int_0^L \Phi_0 \widetilde{c_{sa}}\ \tilde{\theta} 
           (\tilde{v} + v_b).
      \end{equation}
      This is an implicit equation to be solved for the migration speed
      $v_b$, since the Neumann boundary conditions and the right hand side
      of the elliptic equation (\ref{eq:Stokes})) for $\tilde{v}$ (after
      eliminating the pressure $\tilde{p}$) contain expressions that
      depend (linearly) on $v_b$.
\end{enumerate}
Starting with the same unpolarized initial condition as in the
previous Section \ref{sec:spontan}, we obtain a similar
1-dimensional symmetric configuration with no cell translocation and
a central F-actin plateau, a central maximum of FA (focal adhesion)
sites, i.e.~$c_{sa}$-integrins, as well as a
\begin{figure}
  \begin{minipage}[b]{0.5\textwidth}
    \includegraphics[width=1.0\textwidth]{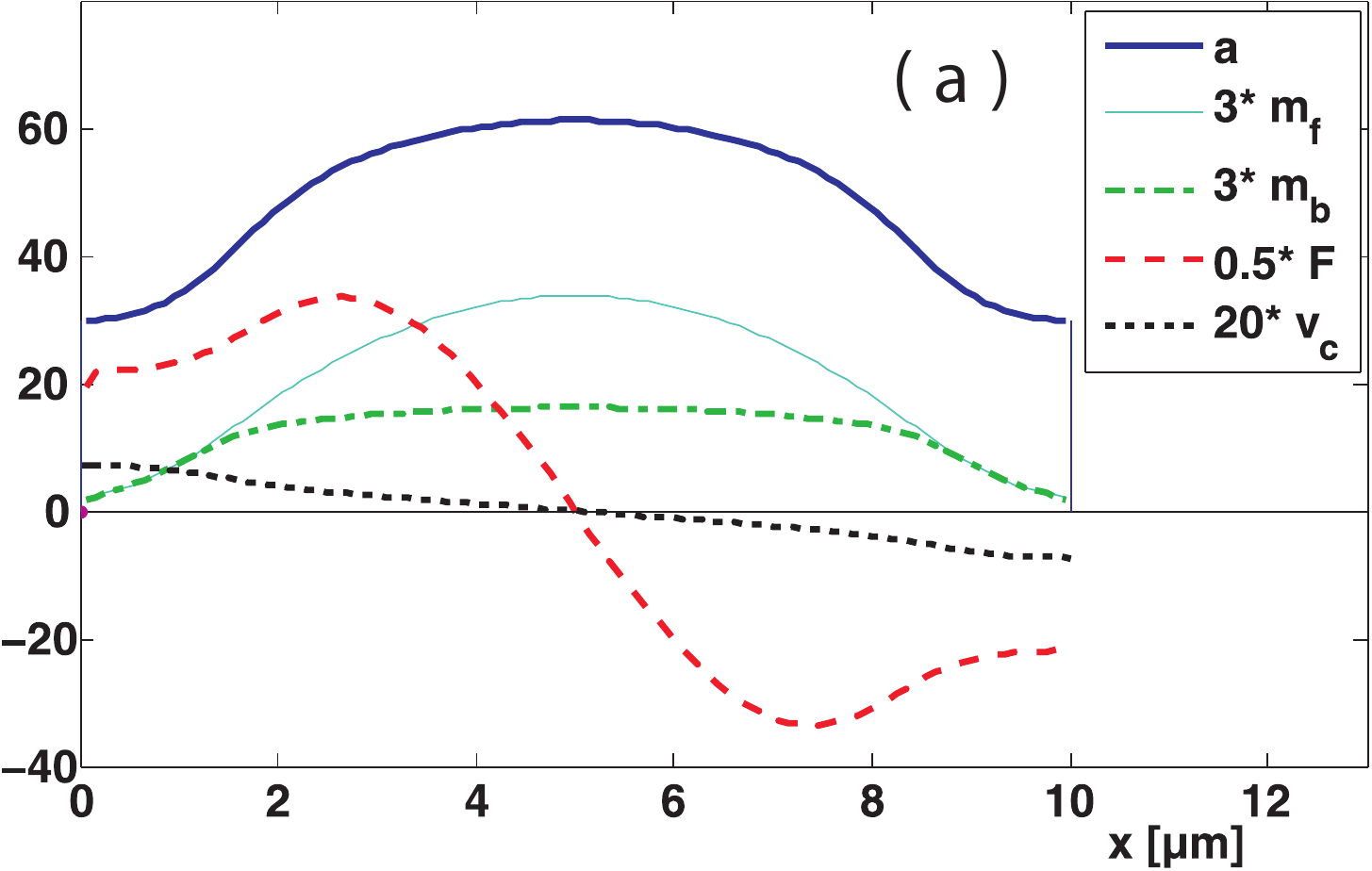} \\
    \includegraphics[width=1.0\textwidth]{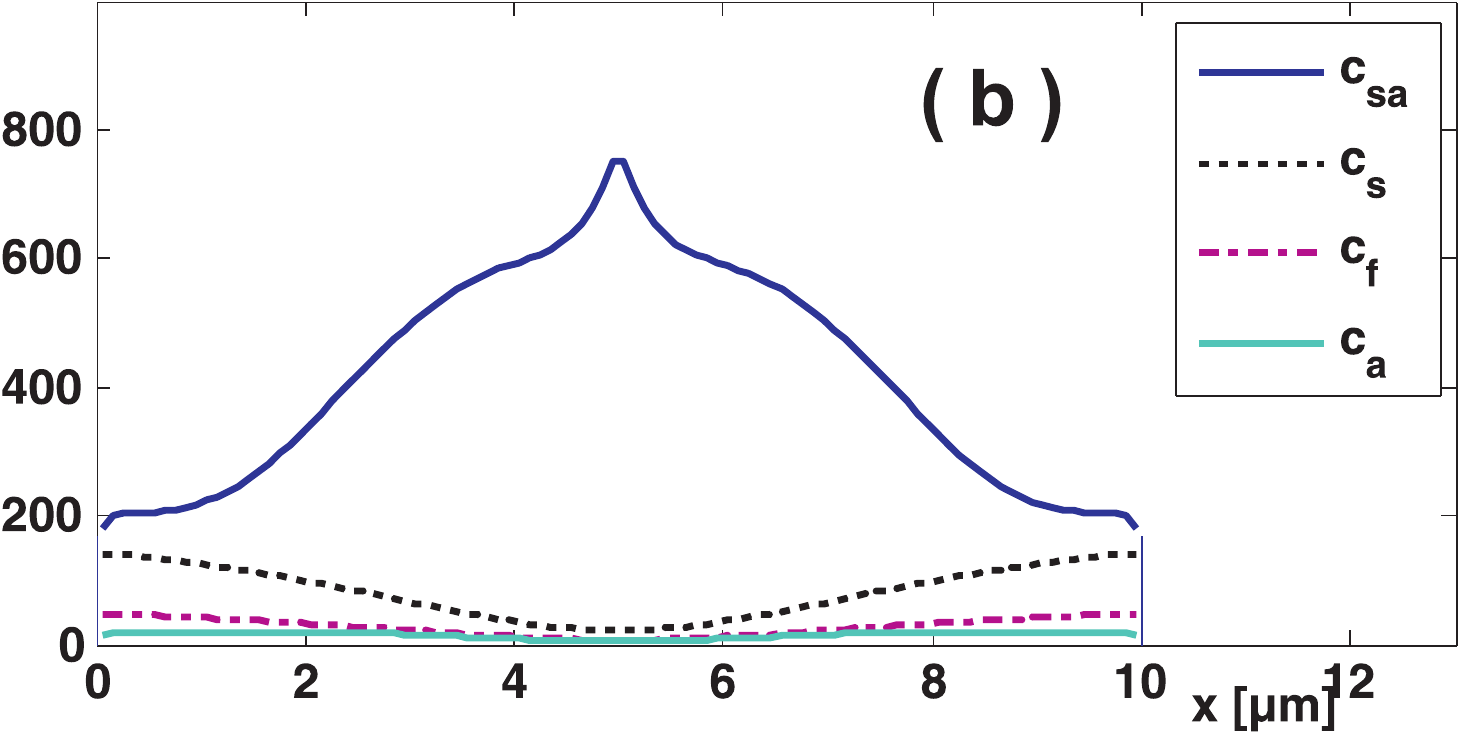}
  \end{minipage} \hskip 2ex
  \begin{minipage}[b]{0.47\textwidth}
    \caption{Symmetric unpolarized state of a model cell (fragment)
             with length $10 \, \mu \mathrm m$
             showing the concentration profiles of \newline
             (a) F-actin  $a$,
                 free and bound myosin-II $m_f$ and
                 $m_b$, the force $F$ onto the substrate
                 and the effective cortical
                 F-actin velocity $v_c$ \newline
             (b) of integrin proteins in the four different states, namely
                 substrate- and actin-bound $c_{sa}$,
                 substrate-bound $c_s$,
                 freely diffusing $c_f$, and
                 actin-bound $c_a$.
            }\label{fig:TauSym}
  \end{minipage}
\end{figure}
centripetal F-actin flow, see Fig.~\ref{fig:TauSym}. In addition,
also the concentration of total myosin-II is enriched in the central
region, with actin-bound myosin forming a central plateau,
consistent with fluorescence pictures of unpolarized keratinocyte
fragments, see \cite{verkhovsky:polarization}. Moreover, the F-actin
flow is highest at the boundary 
(with values $V \sim 0.5 \, \mu \mathrm m /\mathrm{min}$) 
due to the assumed active tip
polymerization.

In contrast to the previous 2-D model simulations, here we find
robust parameter constellations yielding stability of this symmetric
unpolarized state: even quite strong but still subthreshold
perturbations of the F-actin concentration at one side induce only
transient locomotion together with shifts in most concentration
profiles: after some delayed overshooting, the cell fragment returns
to its non-moving non-polarized stable state in
Fig.~\ref{fig:TauSym}. However, if F-actin polymerization is
continuously stimulated at one side (mimicking the effect of a
chemo- or haptotactic gradient) then, not surprisingly, the cell
fragment slowly polarizes and starts to persistently translocate in
this direction (data not shown).

Similarly, in order to mimic the mechanical stimulation experiments
with keratinocyte fragments as performed in
\cite{verkhovsky:polarization}, we locally increase the activation
(and F-actin-binding) of myosin-II at the left hand side for a
certain time (up to $30\sec$), which is thought to be analogous to
the experimental push by a micropipette flow pulse, since by local
compression of the cytoskeleton, filament alignment and thus myosin
action will be enhanced. If time span or amplitude of this
myosin-pulse stays subthreshold, the model cell responses only by a
transient migration as described above and asymptotically returns to
its stable resting state, see the speed curve in
\begin{figure}[htbp]
  \centering
  \includegraphics[height=0.26\textwidth]{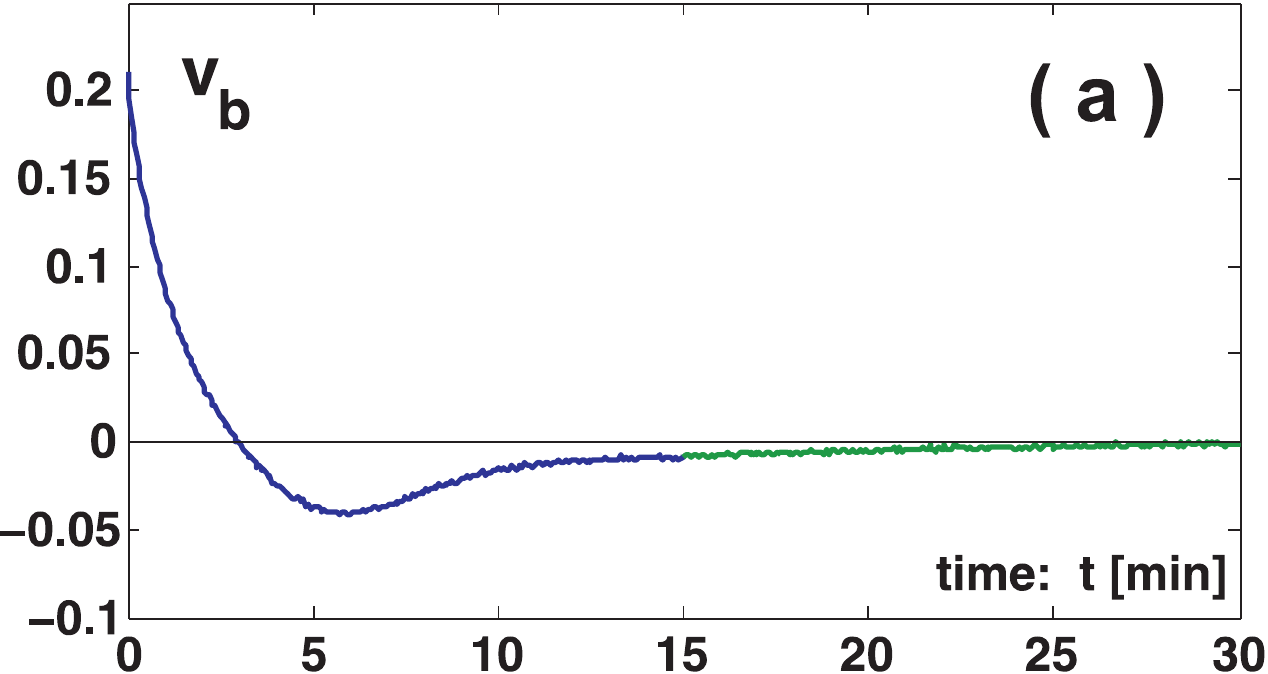} \hskip 2ex
  \includegraphics[height=0.26\textwidth]{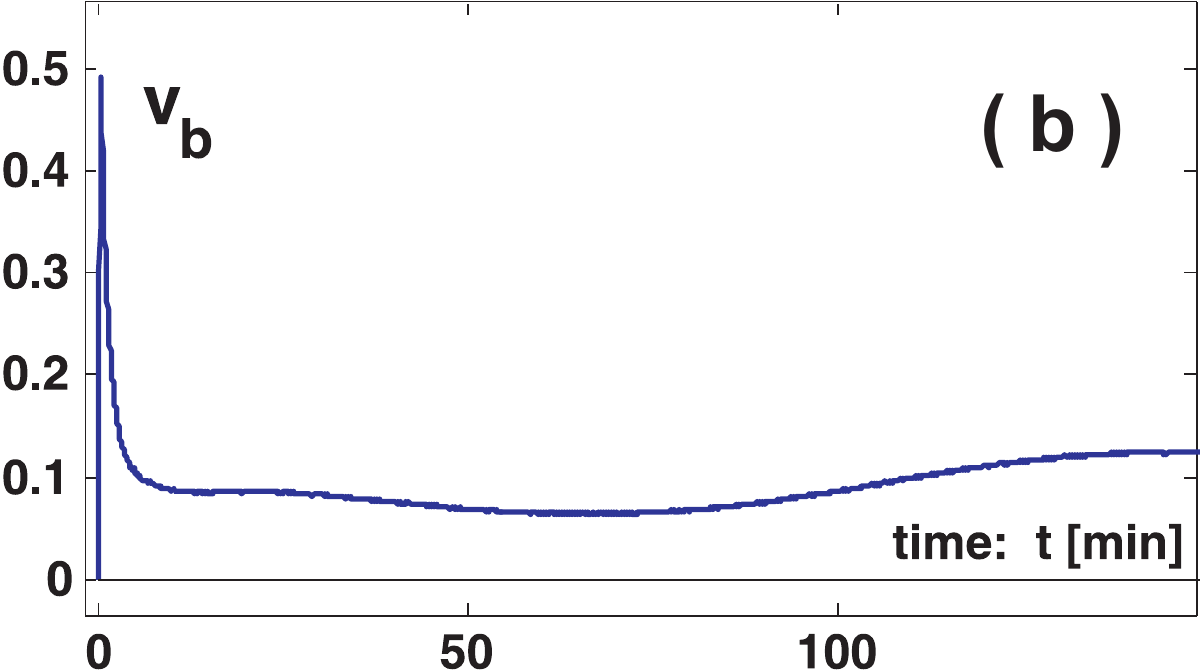}
  \caption{Plots of cell migration velocity $v_b$
           over time after a perturbation
           of the unpolarized state in Fig.~\ref{fig:TauSym} by
           imposing an additional activation rate $\alpha_m^+$
           for actin-bound myosin-II oligomers over an induction
           period of $0.5 \min$ locally at the left cell side:
           (a) with rate $\alpha_m^+=20/\mathrm{min}$, showing a stable
               convergence of $v_b$ towards the unpolarized speed
               zero, and
           (b) with rate $\alpha_m^+=50/\mathrm{min}$, showing the convergence
               towards a positive polarized speed 
               $\sim 0.012 \, \mu \mathrm m / \mathrm{min}$.
  }\label{fig:PolStab}
\end{figure}
Fig.~\ref{fig:PolStab}(a). However, if the pulse strength exceeds a
certain threshold, myosin-II and F-actin is condensed at the rear,
whence active tip polymerization ($V$) is drastically reduced at
this side, so that the cell rapidly starts to migrate in the other
direction, see Fig.~\ref{fig:PolStab}(b). After cessation of the
pulse, the migration speed is reduced, but the cell maintains its
polarized locomotion state. This induced polarization, as a
non-linear threshold behavior, is supported by a further positive
feedback mechanism: Due to cell translocation, the focal adhesion
sites ($c_{sa}$) are successively shifted rearwards relative to the
cell, which stabilizes the asymmetric polarization and later
increases the locomotion speed to a constant asymptotic value 
$(v_b \sim 0.12 \, \mu \mathrm m / \mathrm{min})$.

\begin{figure}[htbp]
  \begin{minipage}[b]{0.5\textwidth}
    \includegraphics[width=1.0\textwidth]{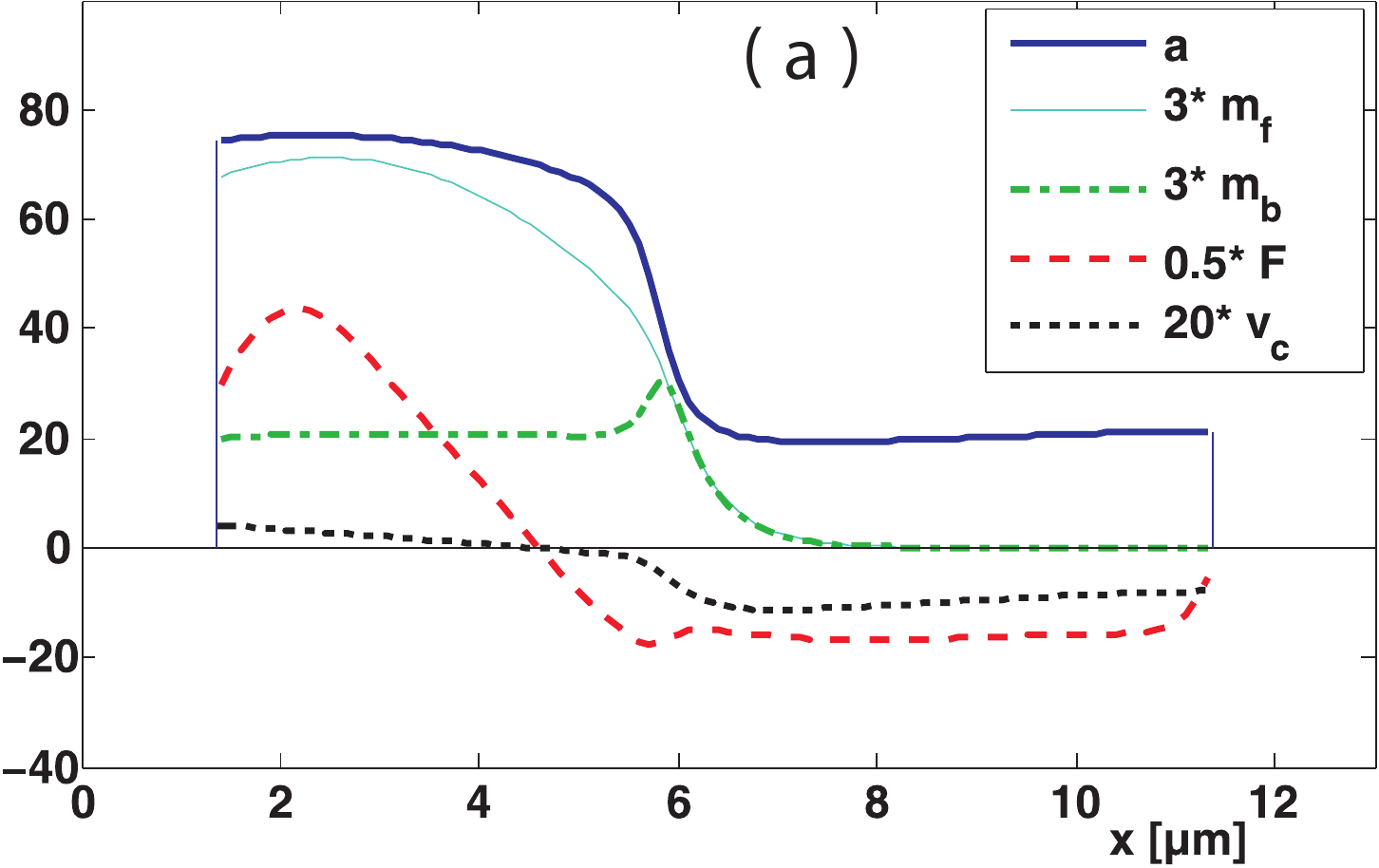}
    \includegraphics[width=1.0\textwidth]{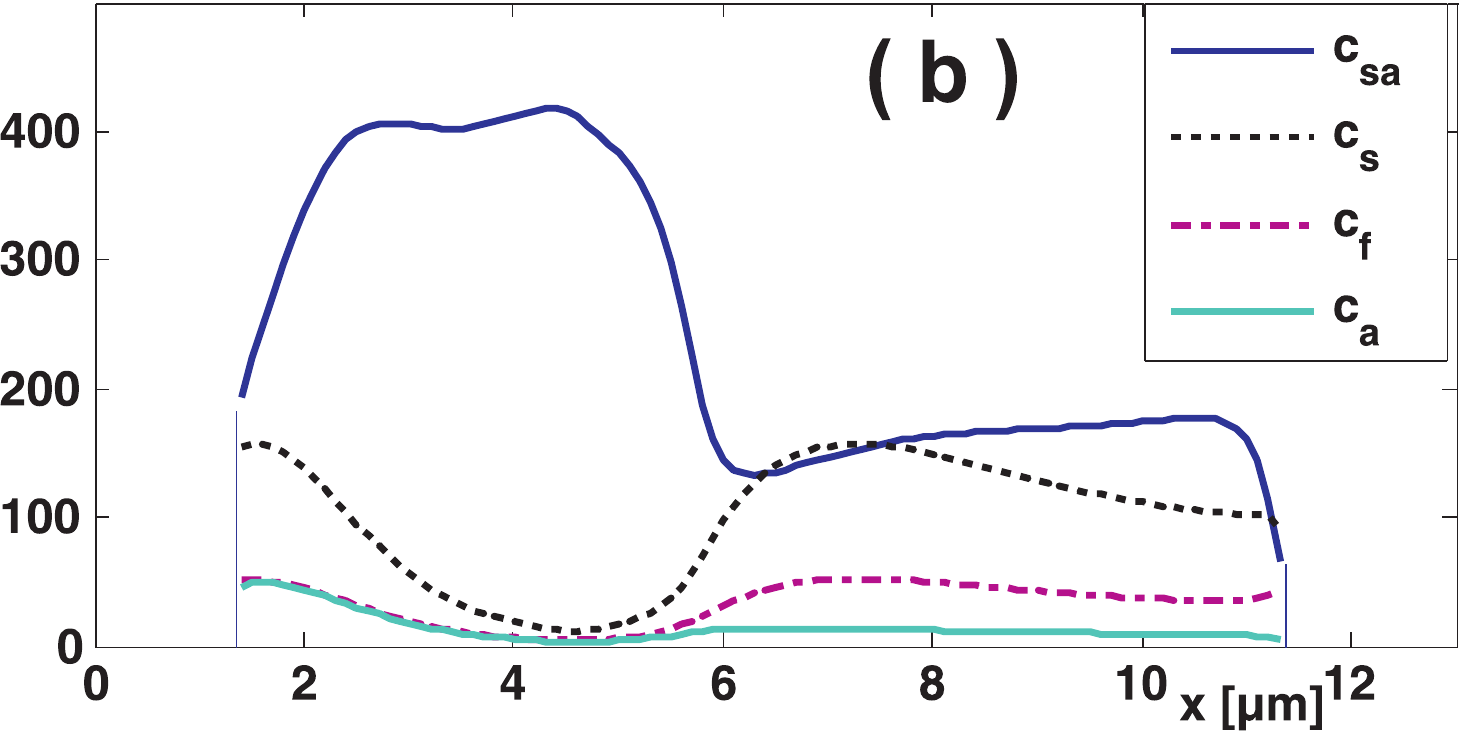}
  \end{minipage} \hskip 2ex
  \begin{minipage}[b]{0.47\textwidth}
    \caption{Migrating polarized state showing
             stable concentration profiles \newline
             (a) of F-actin, $a$, free and bound myosin-II, $m_f$ and
                 $m_b$, the transduced locomotion force $F$
                 and the effective cortical F-Actin velocity
                 $v_c$;
                 \newline
             (b) of integrin-concentrations
                 as in Fig.~\ref{fig:TauSym}.
                 Only FA integrins in the $sa$ state are able to transduce force
                 from the contractile actin network to the substrate.
                 \newline
            The migration speed is 
            $v_b = 0.125 \, \mu \mathrm m / \mathrm{min}$.
    }\label{fig:TauPol}
  \end{minipage}
\end{figure}
In this migration state the cell (fragment) attains characteristic
concentration profiles, see Fig.~\ref{fig:TauPol}. Besides the
already mentioned polar gradients of F-actin and myosin-II, the most
impressive distribution is that of the FA sites: the $c_{sa}$
profile shows a characteristic broad peak behind the leading edge
(as in the 2-D case above), followed by a slight decrease and a
second plateau of even more condensed adhesion sites in the back
part of the cell (fragment) which, however, rapidly decays at the
very rear, see Fig.~\ref{fig:TauPol}(b). This last phenomenon is the
theoretically expected and experimentally observed \emph{rear
release} of adhesion sites or integrins, not induced by any directed
regulatory protein, but only by the fact that the rear part
experiences a steep increase of the force $F$ transduced to
the substratum, see the plot in Fig.~\ref{fig:TauPol}(a). By
(\ref{eq:AdhesionForce}) this is proportional to the F-actin mass
flow $\theta v_c$ with respect to the substratum: While in
the major front part of the migrating cell (fragment) the F-actin
flow is retrograde and the centripetally pulling negative force is modest in
amplitude, near the trailing edge the direction of flow reverses and
the positive force becomes very strong, now centripetally pulling off the
focal adhesion sites. Thus, the reason for cell translocation is not
that there is ``more adhesion'' or ``stronger force'' at the front
compared to the rear, as it is asserted in some models,
cf.~\cite{DiMillaLauffenburger}. Indeed, for vanishing passive
friction $\Phi_u$ due to (\ref{eq:SubstrateForce}) the total force
integral even vanishes. The true physical reason for cell migration
rather is the aforementioned asymmetry in the polarized cell state,
expressed by a wide front region with modest rearward force and
short rear region with strong forward force.

We finally investigate the dependence of migration speed on two cell
physiologically important parameters, namely the \emph{adhesiveness
of the substratum} quantified by the relative number of available
adhesion sites $\mathrm{Adh0}$, and the \emph{responsiveness of the
F-actin network} measured, for instance, by the concentration
$\mathrm{Arp0}$ of activated Arp2/3 proteins. These are mainly
responsible for controlling F-actin polymerization by available free
filament ends as they appear in the net assembly rate
(\ref{eq:Assembly}) and the ratchet polymerization pressure
\begin{figure}[htbp]
  \includegraphics[width=0.495\textwidth]{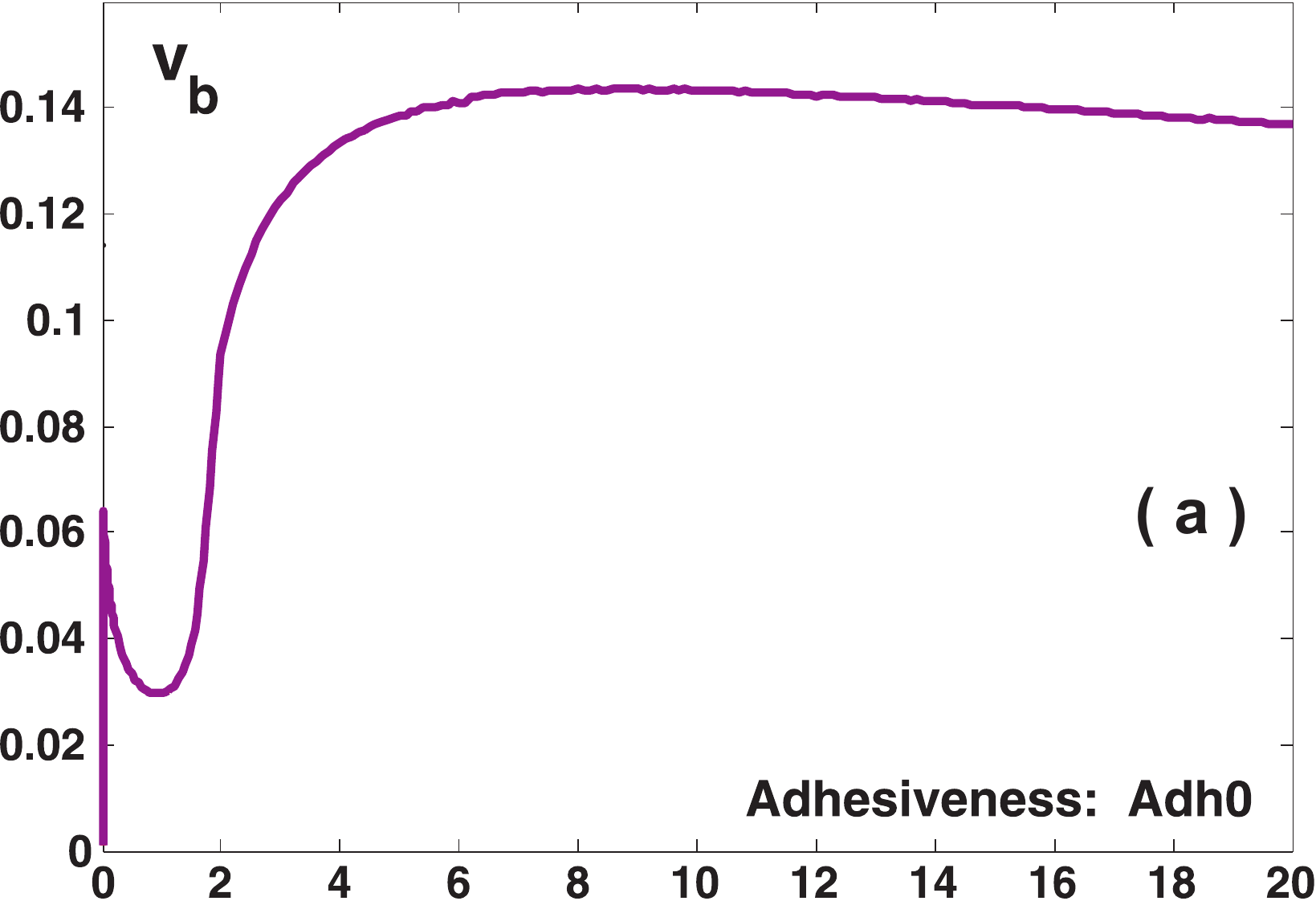}
  \includegraphics[width=0.495\textwidth]{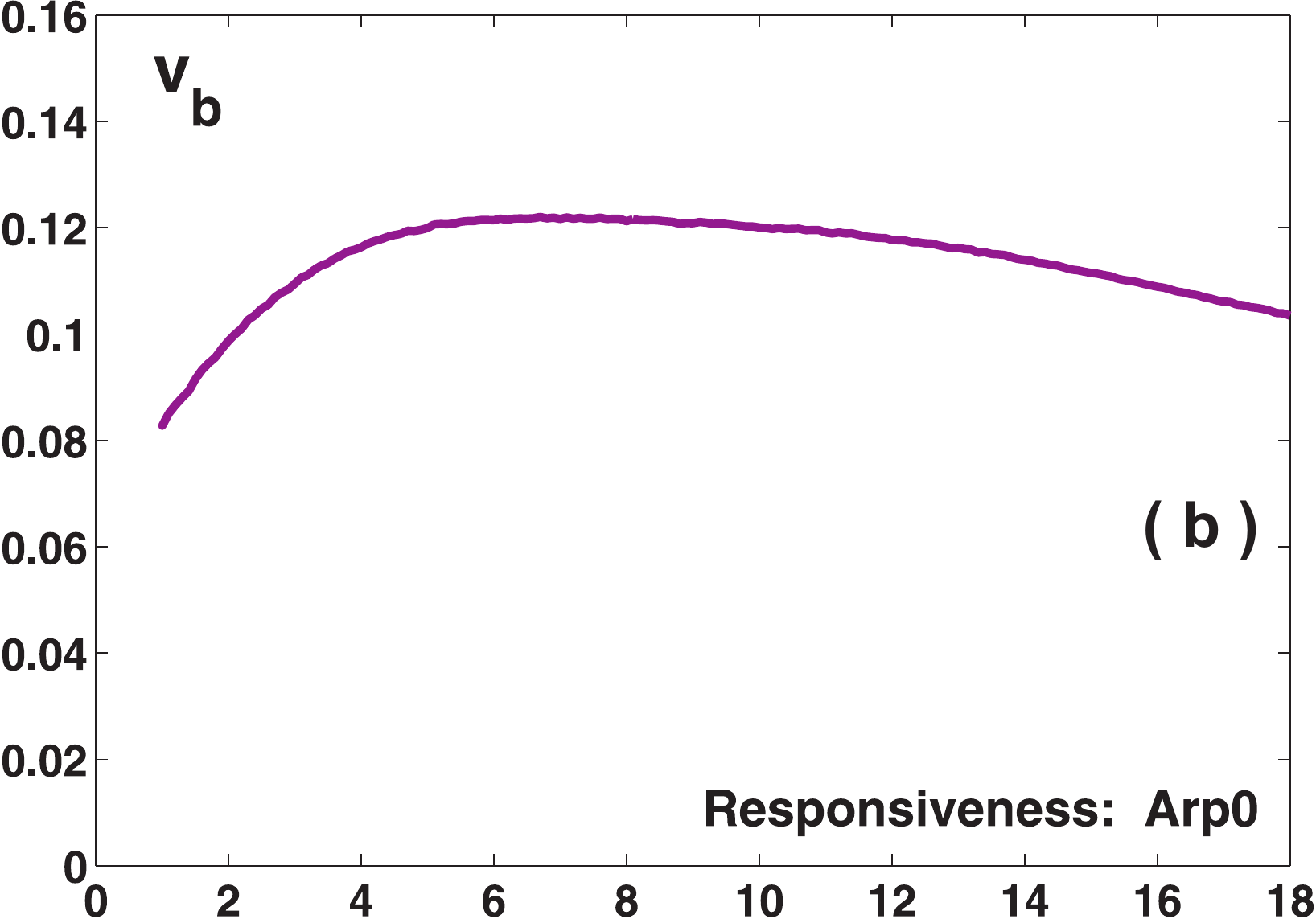}
  \caption{Migration speed $v_b$ of a model cell in its stable
           polarized state, plotted over varied parameters 
           (a) of \emph{adhesiveness} $\mathrm{Adh0}$, 
               relative amount of adhesion
               binding sites on the substratum (e.g.~fibronectin
               coating) 
           (b) of \emph{F-actin responsiveness} expressed by $\mathrm{Arp0}$
               $[\mu \mathrm M]$, the
               cytoplasmic concentration of activated Arp2/3 complexes.
          }\label{fig:CurvesMax} 
\end{figure}
(\ref{eq:Ratchet}). The results depicted in Fig.~\ref{fig:CurvesMax}
reveal the existence of optimal ranges for both parameters,
consistent with earlier modeling results, see
\cite{DiMillaLauffenburger,grachevaothmer}, and with experimental
observations, see e.g.~\cite{hinz,Palecek97}.

Particularly, the migration response curve of
Fig.~\ref{fig:CurvesMax}(a) has been performed for fixed parameter
$\mathrm{Arp0}=10 \, \mu \mathrm M$ and for adhesion values $0 \leq
\mathrm{Adh0} \leq 20$, where according to Table~\ref{tab:parameter}
the adhesiveness proportionally influences not only the two adhesion
rates $\alpha^+ = \gamma^+ = \mathrm{Adh0}\cdot\alpha$ but also the
passive membrane friction coefficient $\Phi_u = \mathrm{Adh0} \cdot
6 \, \mathrm{Pa} \cdot \mathrm{min} \cdot \mu \mathrm m^{-2} $. This
is based on the idealizing assumption that there are no other
relevant exterior forces that would resist locomotion relative to
the substratum, than those due to friction of the dorsal membrane
with respect to an adhesive coating of e.g.~fibronectin or collagen.

Under these hypotheses and parameter choices our model simulations
predict a minimal migration speed of $v_b \sim 0.035 \, \mu \mathrm m
/ \mathrm{min}$ for $\mathrm{Adh0} \sim 1$, which increases towards
the doubled speed when lowering the adhesiveness to almost zero.
This surprising phenomenon is consistent with own experimental
measurements of human keratinocyte polarization and migration
\cite{LibotteBretschneider} revealing slightly increased motility of
polarized cells on glass compared to those on a low density
fibronectin coat. The reason for this can be seen from the
corresponding profiles of FA concentration ($c_{sa}$) and substrate
force distribution ($F$) for the two adhesiveness values
\begin{figure}[htbp]
  \centering
  \includegraphics[width=0.49\textwidth]{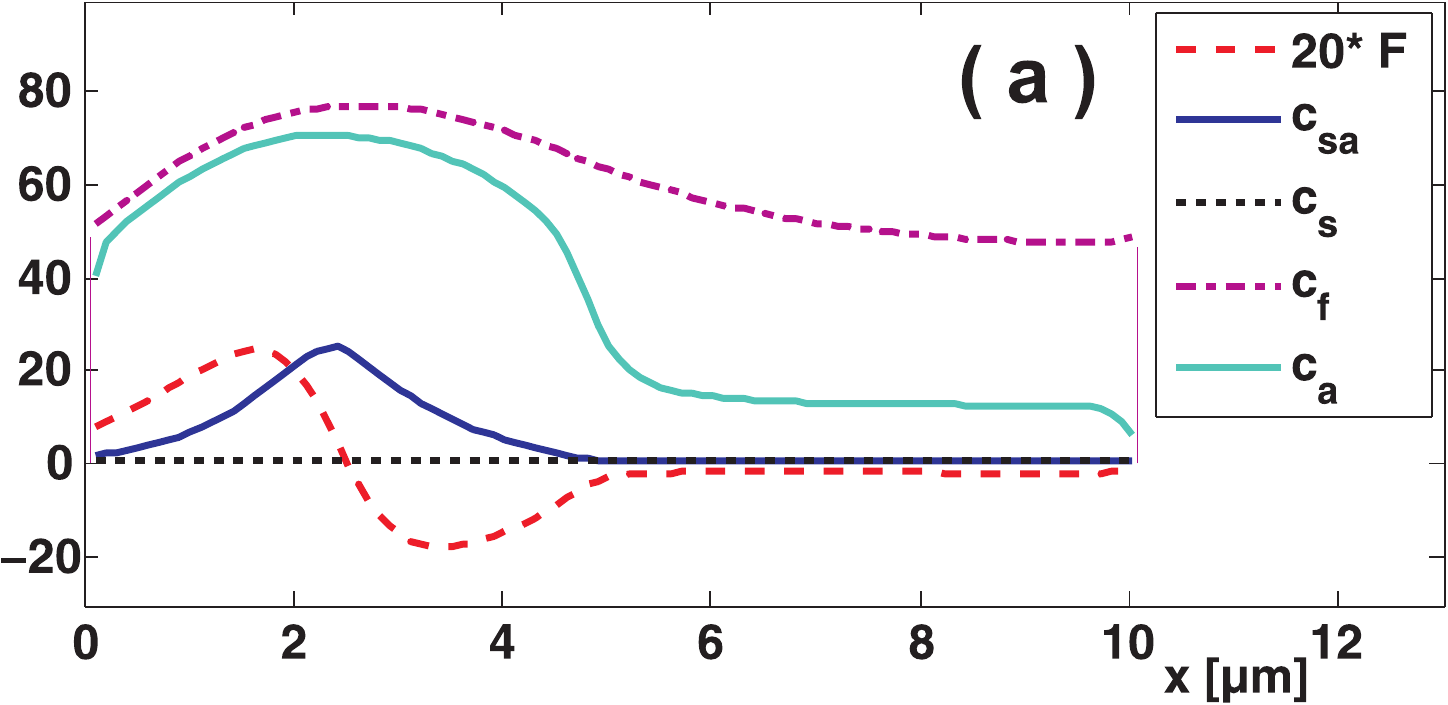} \hskip 1ex
  \includegraphics[width=0.49\textwidth]{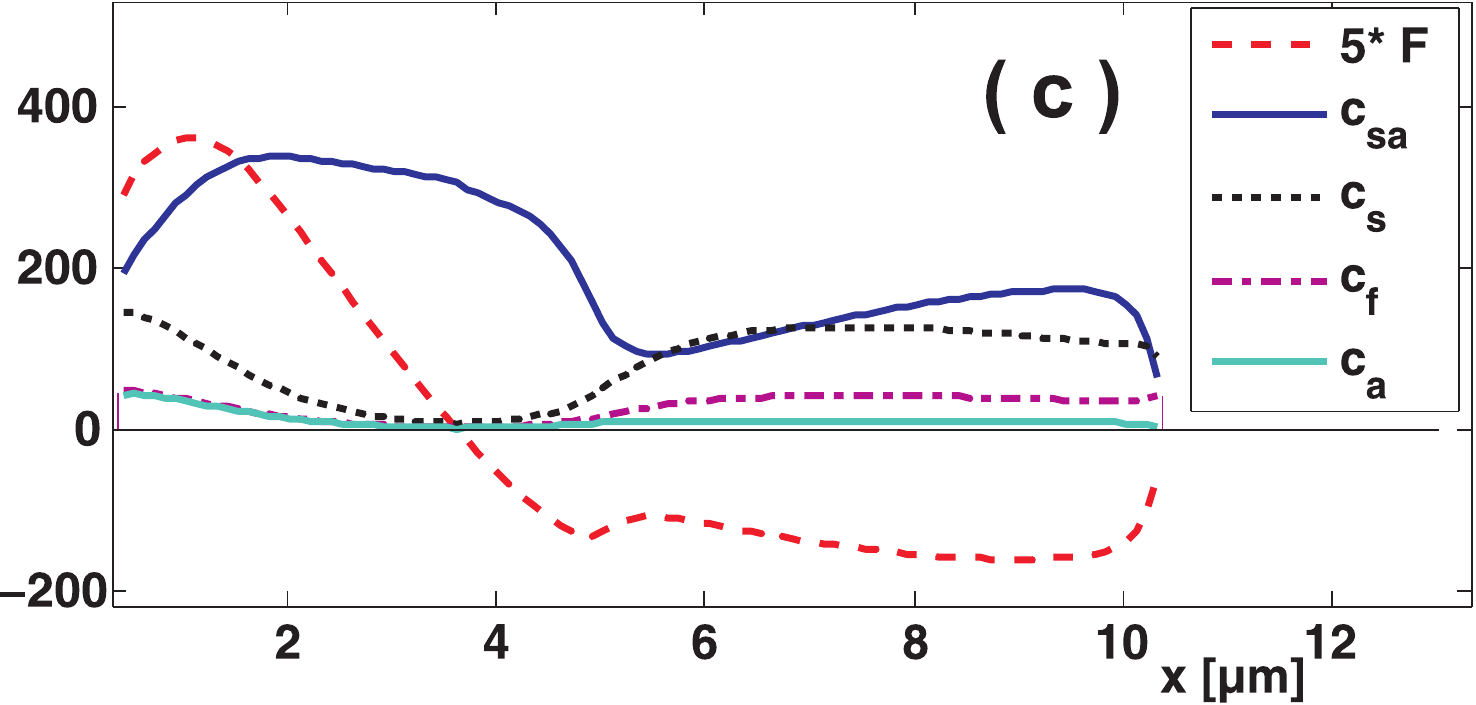} \\
  \includegraphics[width=0.49\textwidth]{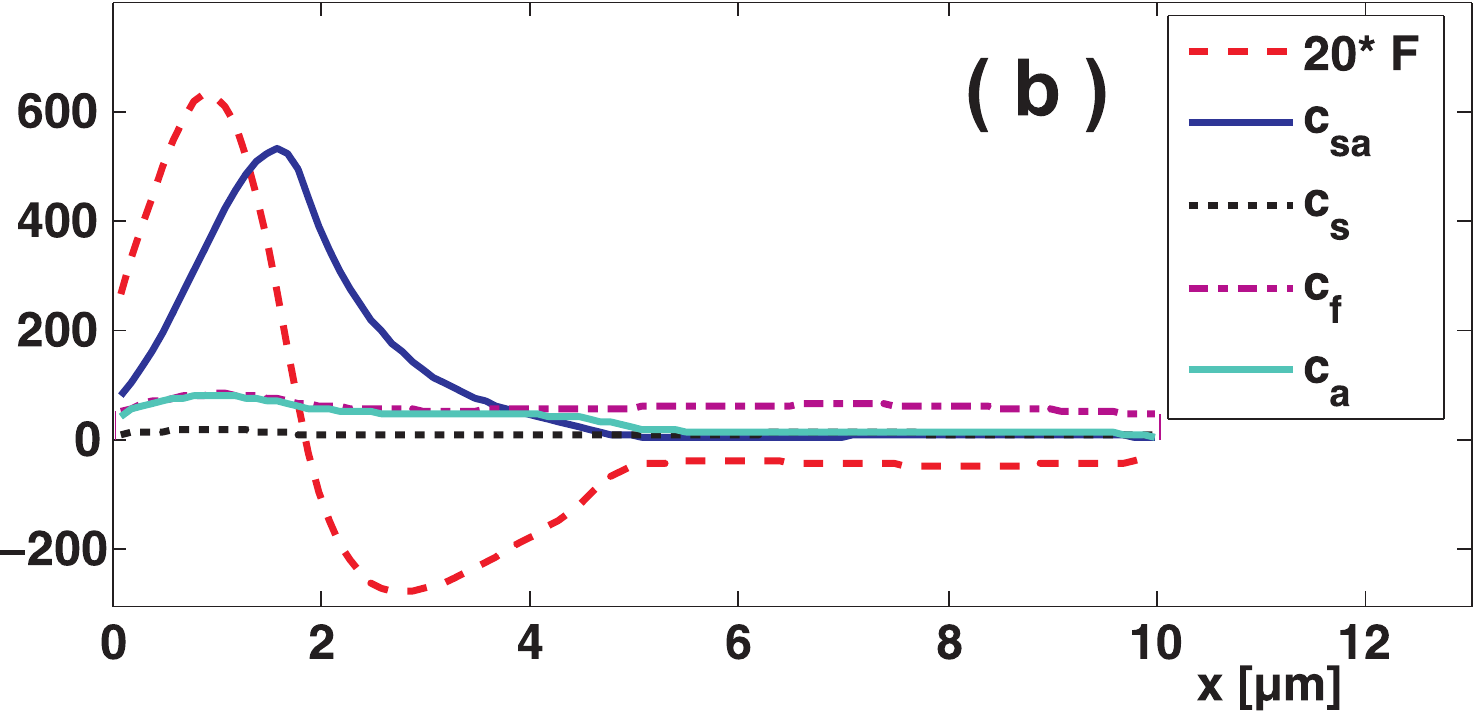} \hskip 1ex
  \includegraphics[width=0.49\textwidth]{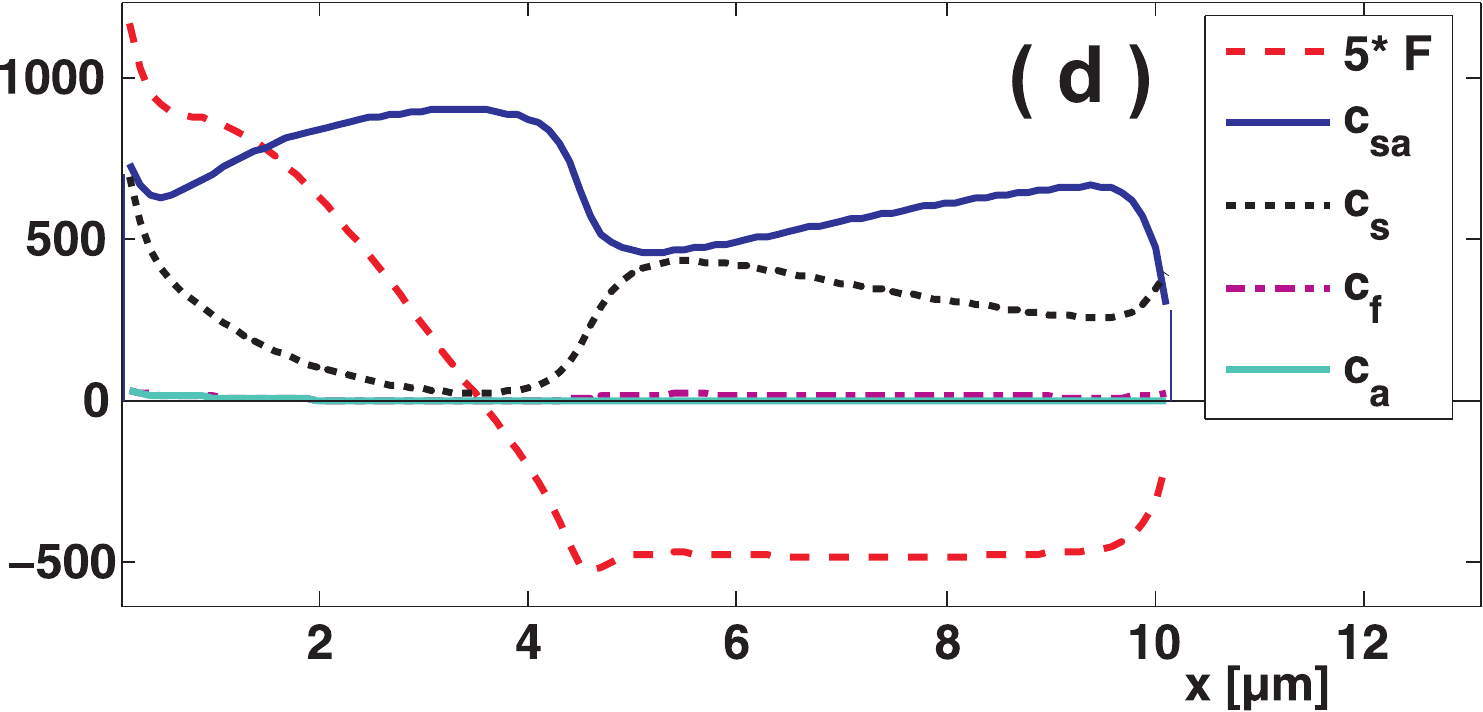}
  \caption{Substrate force distribution $F$
           and integrin concentrations
           in the four different states, plotted for varying
           adhesiveness parameter $\mathrm{Adh0}$
           with corresponding migration
           speed $v_b$ as in Fig.~\ref{fig:CurvesMax}(a):
           (a) $0.01$, speed $0.06 \, \mu \mathrm m / \mathrm{min}$, 
           (b) $0.2$,  speed $0.04 \, \mu \mathrm m / \mathrm{min}$,
           (c) $3.0$,  speed $0.125 \, \mu \mathrm m / \mathrm{min}$, and 
           (d) $20.0$, speed $0.138 \, \mu \mathrm m / \mathrm{min}$.
  }\label{fig:integrinsasfsa}
\end{figure}
$\mathrm{Adh0}=0.01, 0.2$, see Fig.~\ref{fig:integrinsasfsa}(a,b).
Due to ongoing retrograde flow of F-actin, working against internal
viscosity and drag, the polarized cell (fragment) gathers the few FA
sites on the back side in a way that the local FA and force
distributions are almost symmetric (for almost vanishing
adhesiveness $\mathrm{Adh0}=0.01$) but still with an additional
negative (pulling) force plateau on the front side (though of tiny
absolute value $|F| \sim 0.05 \, Pa$). For increased but still
small adhesiveness ($\mathrm{Adh0}=0.2$), the much more frequent FA
sites start to accumulate at the rear, thus the asymmetry of forces is
enhanced and the increased friction reduces the migration
speed.

On the other hand, for further increasing adhesiveness,
$\mathrm{Adh0}>1$, the front plateau of pulling forces at the
enriched FA `carpet' is proportionally increased, see
Fig.~\ref{fig:integrinsasfsa}(c,d). However, the dominant reason for
the non-linear speeding-up response is the prominent increase of
disruptive forces $|F_c(\text{rear})|$ up to values of
$\mathrm{Adh0} \sim 2$, leading to a drastic reduction of resistive
FA sites at the very rear, see the corresponding indicator curves in
\begin{figure}[htbp]
  \includegraphics[width=0.49\textwidth]{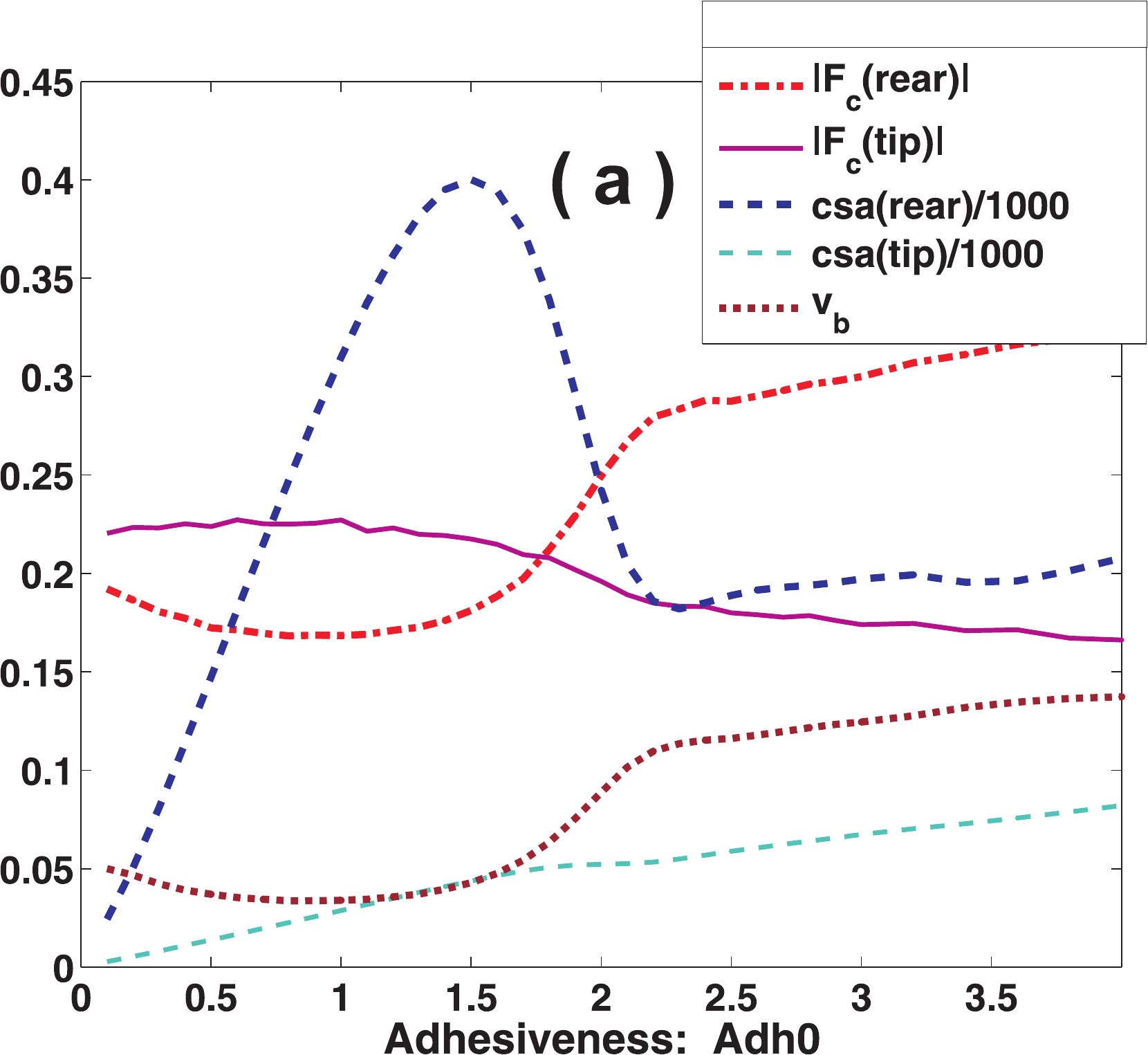}
  \hskip 1 ex
  \includegraphics[width=0.49\textwidth]{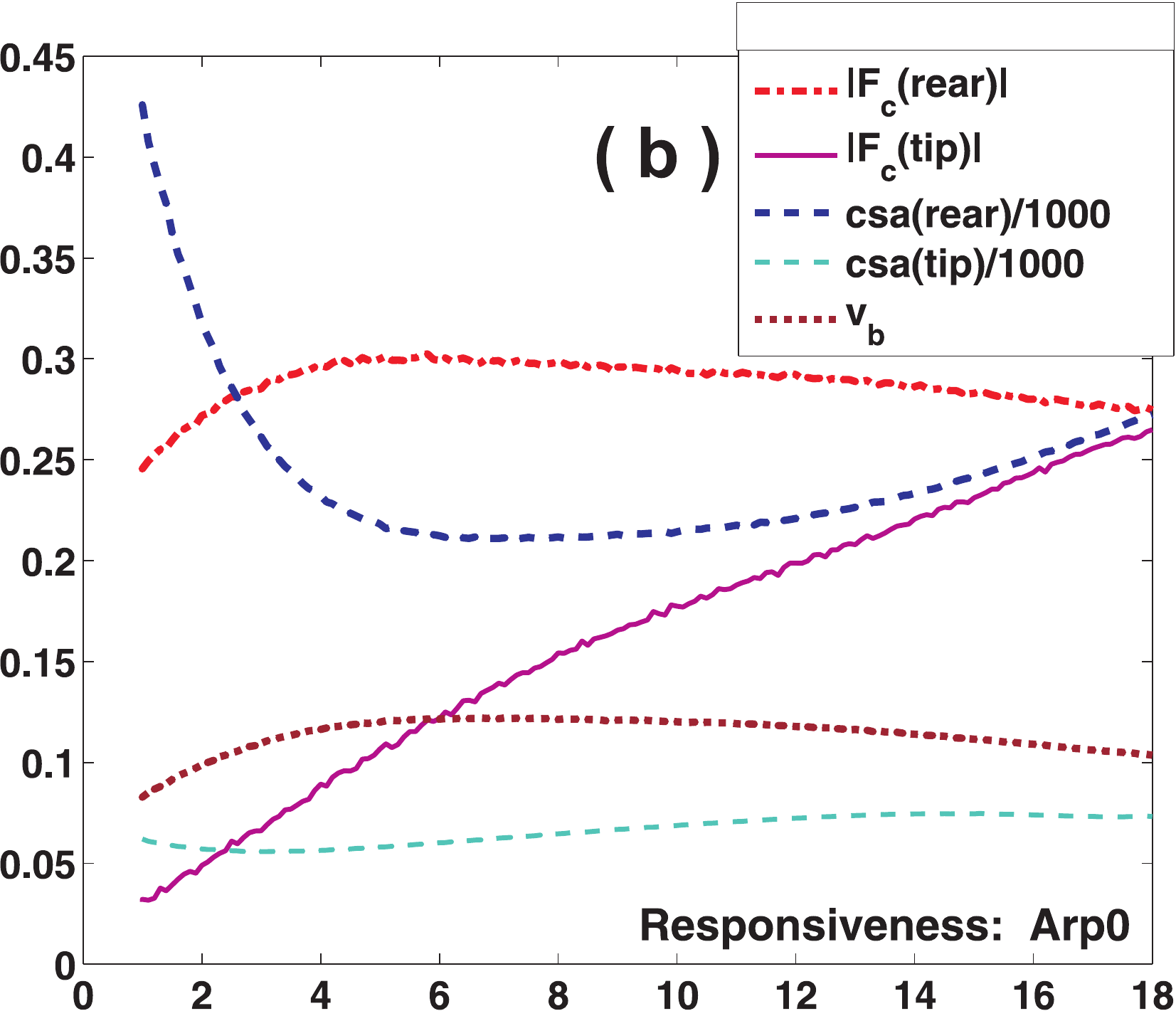}
  \caption{Variation of migration speed $v_b$ and certain indicator
           functions at the cell edge in dependence of
           (a) adhesiveness $\mathrm{Adh0}$ and
           (b) F-actin responsiveness $\mathrm{Arp0}$ $[\mu \mathrm M]$:
           Plotted are the load $|F_c|$ per FA site at the cell
           rear (bold) and tip (light), and the FA concentration
           $c_{sa}$ at the rear (bold, dashed) and tip (light, dashed).
           Ripples in the force curves reflect the slight stochastic
           noise that was added in the force equation (\ref{eq:Stokes}).
          }\label{fig:CurvesFcsa}
\end{figure}
Fig.~\ref{fig:CurvesFcsa}(a). Only for very large adhesiveness,
$\mathrm{Adh0} \sim 20$, the FA sites again start to accumulate at
the very rear in spite of a strong disruptive force there, as can be
seen in Fig.~\ref{fig:integrinsasfsa}(d), causing a slight
decrease in migration speed. Thus, while an adhesiveness between $5$
and $12$ leads to a saturated optimality in migration speed,
within the range $1<\mathrm{Adh0}<4$ the cell (fragment) has a high
sensitivity for responding to an increase of adhesiveness with an up
to 4-fold increase of speed. Carried over to the mean forward speed
of competing leading lamellae in a whole cell model (as for the 2-D
simulations in Fig.~\ref{fig:cellFragment}), this could be used to
explain polarization and haptotaxis of cells in spatial adhesion
gradients, cf.~\cite{frank:keratinocytepolarization}.

\begin{table}[htbp]
\centering 
\normalsize{
\begin{tabular}{lll}
\textbf{Symbol} & \textbf{Meaning} & \textbf{Units} \\ \hline
$\theta(t,x)$ & Volume fraction of F-actin            & dimensionless \\
$a(t,x)$   & Concentration of F-actin                 & $\mu \mathrm M$ \\
$m_f(t,x)$ & Concentration of free myosin-II          & $\mu \mathrm M$ \\
$m_b(t,x)$ & Concentration of F-actin bound myosin-II & $\mu \mathrm M$ \\
$c_f(t,x)$ & Concentration of free integrin           & $\mu \mathrm m^{-2}$ \\
$c_s(t,x)$ & Concentration of substrate bound integrin & $\mu \mathrm m^{-2}$ \\
$c_a(t,x)$ & Concentration of actin bound integrin    & $\mu \mathrm m^{-2}$ \\
$c_{sa}(t,x)$ & Concentration of substrate-and-actin bound integrin
                                                     & $\mu \mathrm m^{-2}$ \\
$R(\theta)$        & F-actin net polymerization rate & $\mathrm{min}^{-1}$ \\
$\psi(\theta,m_b)$ & Contractile stress              & $\mathrm{Pa}$ \\
$\sigma(\theta)$   & Swelling pressure               & $\mathrm{Pa}$ \\
$S(\theta,m_b)$    & Effective cytoplasmic stress    & $\mathrm{Pa}$ \\
$p(t,x)$           & Effective two-phase flow pressure & $\mathrm{Pa}$ \\
$\tau_u(t,x)$      & Dorsal membrane tension         & $\mathrm{Pa}$ \\
$\mathbf v(t,x)$   & Mean F-actin velocity  & $\mu \mathrm m/ \mathrm{min}$ \\
$\mathbf v_c(t,x)$ & Cortical F-actin velocity
                                            & $\mu \mathrm m / \mathrm{min}$ \\
$\mathbf u(t,x)$   & Dorsal membrane velocity & $\mu \mathrm m/ \mathrm{min}$ \\
$v_b(t)$      & Migration velocity of cell (fragment) body (1-D)
                                            & $\mu \mathrm m / \mathrm{min}$ \\
$\Phi(\theta,c_{sa})$ & Adhesional friction & $\mathrm{Pa}\cdot\mathrm{min} / \mu \mathrm m ^2$ \\
$\mathbf F_v$ & Active frictional force     & $\mathrm{Pa} / \mu \mathrm m $\\
$\mathbf F_c$ & Local frictional force per adhesion site
                                            & $\mathrm{Pa} \cdot \mu \mathrm m $\\
$\mathbf F_u$ & Passive frictional force at dorsal membrane
                                          & $\mathrm{Pa} / \mu \mathrm m $ \\
$\dot\Gamma$  & Normal speed of the free boundary $\Gamma(t)$
                                          & $\mu \mathrm m/\mathrm{min}$ \\
$V$           & Relative inward F-actin velocity at boundary
                                          & $\mu \mathrm m/\mathrm{min}$\\
$P_a^\Gamma$  & Boundary cytoskeleton pressure       & $\mathrm{Pa}$\\
$P_s^\Gamma$  & Boundary cytosol pressure            & $\mathrm{Pa}$\\
$\tau_\Gamma$ & Membrane tension at boundary (lamellar tips) & $\mathrm{Pa}$\\
$a^B(a)$      & Concentration of tip-bound F-actin        & $\mu \mathrm M$ \\
$a^c(a)$      & Concentration of F-actin bound to clamp-motor proteins        & $\mu \mathrm M$ \\
$a^f(a)$      & Concentration of free filaments exposed to the tip      & $\mu \mathrm M$ \\
$\mu_\Gamma(m_b)$   & F-actin shear viscosity relative to $a^B$
                        filaments    & $\mathrm{Pa} \cdot \mathrm{min} / \mu \mathrm m  $ \\
$p^f$         & Free polymerization pressure at tip membrane       & $\mathrm{Pa}$\\
$p^c$         & Clamp-motor polymerization pressure        & $\mathrm{Pa}$\\
\hline
\end{tabular}
} \caption{List of model variables and
functions}\label{tab:variablen}
\end{table}

In order to understand the analogous optimal migration performance
for responsiveness parameters in the range between 
$4 \, \mu \mathrm M$ and $10 \, \mu \mathrm M$ 
of Arp2/3 concentration $\mathrm{Arp0}$,
see Fig.~\ref{fig:CurvesMax}(b), we plot the same indicator curves
for FA concentrations and substrate forces in
Fig.~\ref{fig:CurvesFcsa}(b). Again, with increasing Arp2/3-induced
actin polymerization there is an almost linear increase in pulling
force at the tip (due to increased polymerization velocity $V$
there) while the concave migration speed curve directly corresponds
to the proportional curve for the disruptive forces at the rear and
the resulting convex curve for the accumulated FA sites there. For
larger $\mathrm{Arp0}$ values the growing accumulation of FA sites
again leads to a slight speed reduction. Finally, the about $50\%$
gain of speed for increasing responsiveness in the range 
$1 \, \mu \mathrm M < \mathrm{Arp0} < 4 \, \mu \mathrm M$ 
could again be carried
over to a competing lamellar protrusion response in spatial
chemotactic gradients, which are known to stimulate F-actin
polymerization at the leading front.

\subsection{Migration speed in a simplified 1-D model}

In order to explore, which properties of our coupled
cytoplasm-adhesion model are essential for obtaining the 1-D
simulation results in the preceding section, we simplified the model
by freezing the following variables and parameters,
cf.~\cite{christoph:dipl}:
\begin{enumerate}
  \item Assuming fast diffusion and F-actin binding of free myosin-II
        oligomers, we take the pseudo-steady state condition
        (\ref{eq:MyosinStar}) for bound myosin-II, but with the
        equivalent dissociation rate
        $\delta_m(a)= \delta_m^0 \exp(-2a/a_\mathrm{opt})$
        instead of (\ref{eq:delta-m}), and obtain as contractility
        $\psi(\theta) = \widetilde{\psi_0} \cdot \theta^2
                         \exp(-2\theta / \theta_\mathrm{opt})$
        with a coefficient $\widetilde{\psi_0}$ analogous to
        (\ref{eq:ContractionTheta}).
  \item Except the FA friction function $\Phi$ we set all other
        friction coefficients ($\varphi$,$\Phi_u$) to zero. Thus, the cell
        migration speed is determined by the zero integral in
        (\ref{eq:SubstrateForce}).
  \item All `exterior' pressures or tensions at the free membrane tip
        ($\pi_0^f$,$\pi_0^c$,$\tau_\Gamma$) are assumed to vanish,
        in particular,
        there occurs no active polymerization ($V \equiv 0$).
\end{enumerate}

\begin{figure}[htbp]
    \centering
    \includegraphics[width=0.445\textwidth]{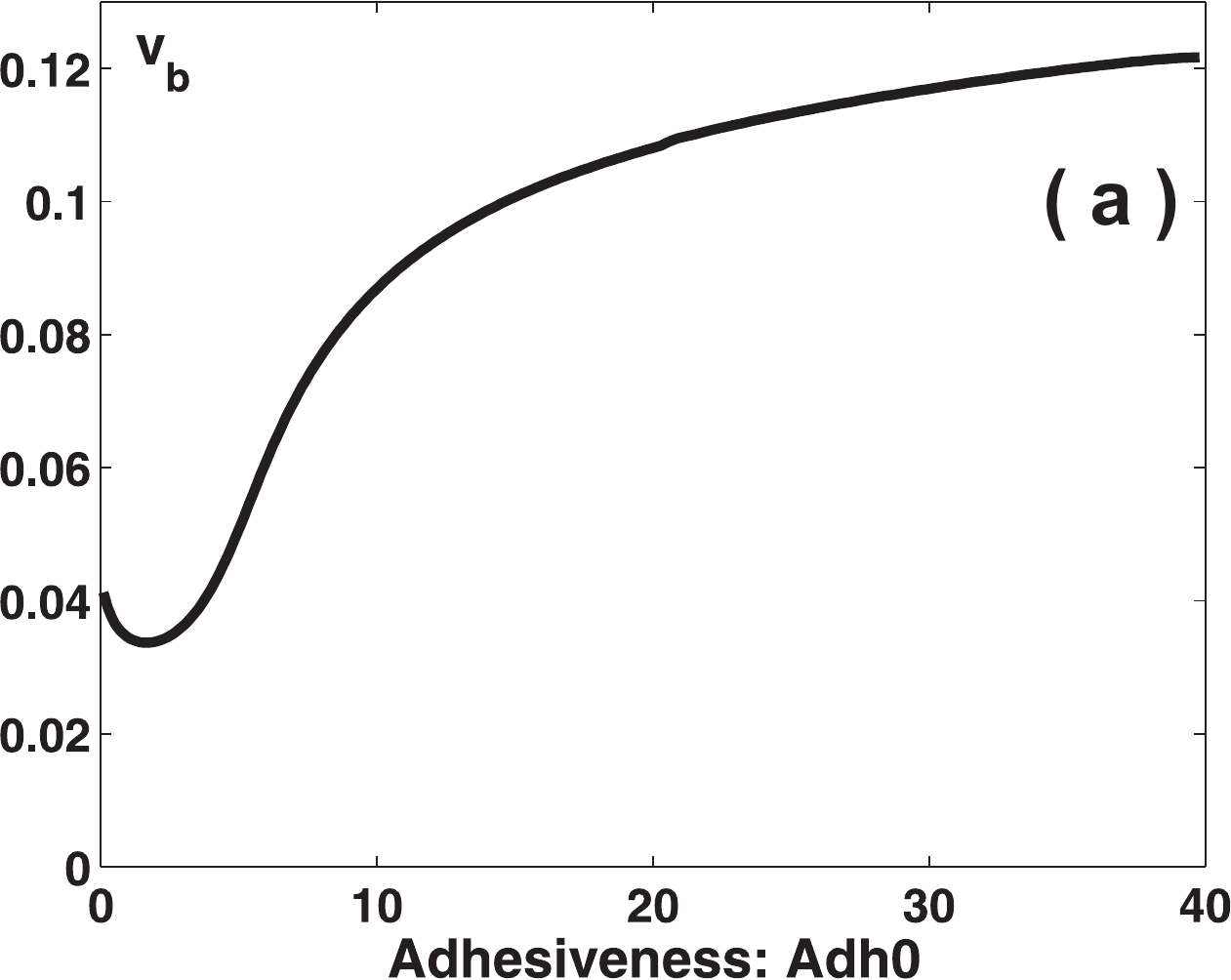} \hskip 2ex
    \includegraphics[width=0.445\textwidth]{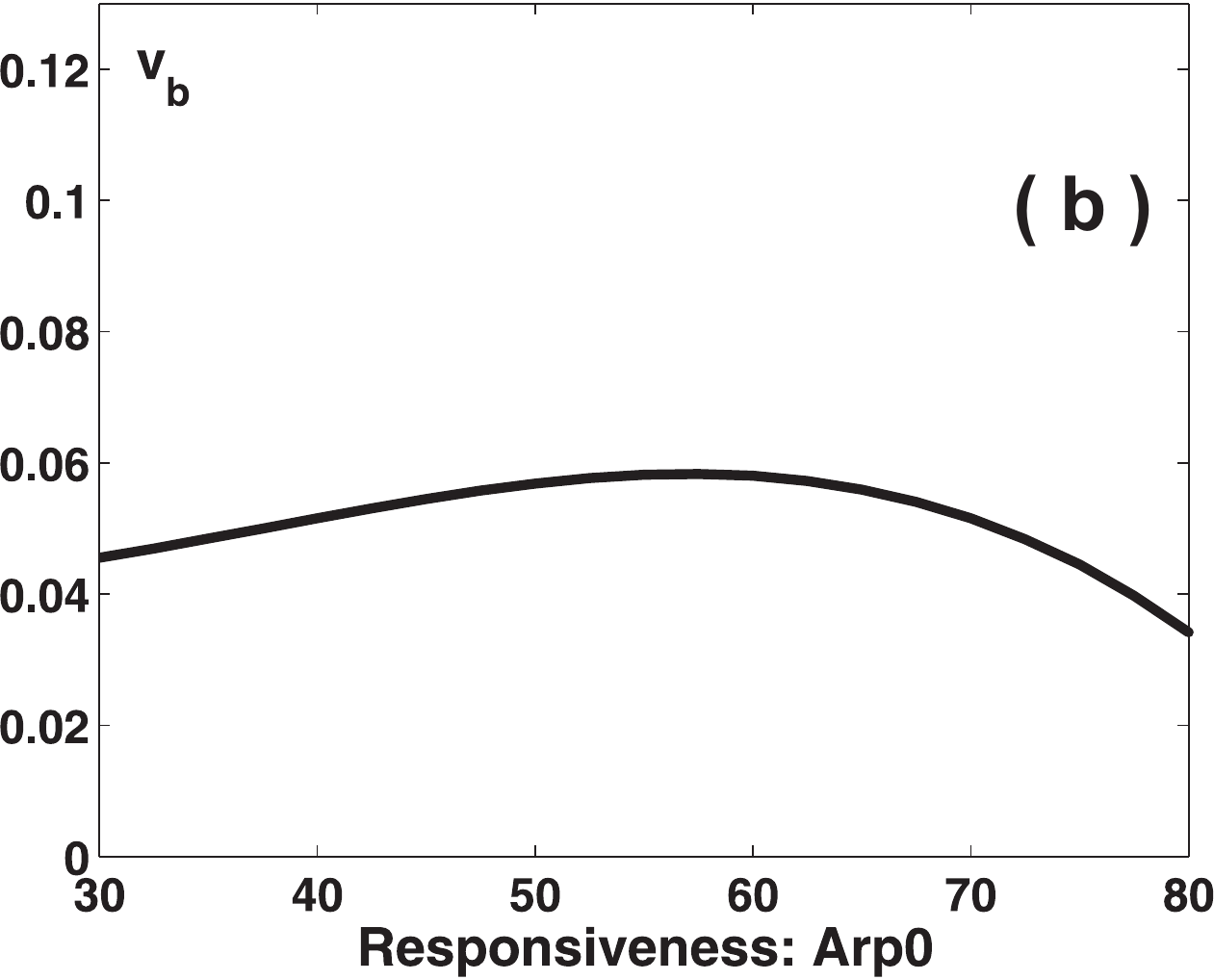}
    \caption{Migration speed $v_b$ for a simplified model cell in its
             stable polarized state in dependence of ($a$) adhesiveness
             or ($b$) F-actin responsiveness as described in
             Fig.~\ref{fig:CurvesMax}.
            }
    \label{fig:integrins5}
\end{figure}

Starting with uniform FA concentration and a minor asymmetry in the
distribution of actin, the system reaches a polarized and migrating
state and the distributions of the various concentrations exhibit
the same characteristics as in the full model (Fig
\ref{fig:TauPol}). We again analyze the effect of substratum
adhesiveness and F-actin responsiveness on the migration speed, see
Fig.~\ref{fig:integrins5}.
In contrast to the observations for the full model (see
Fig.~\ref{fig:CurvesMax}), a saturation behavior of the migration
velocity emerges with increasing adhesiveness instead of an optimal
range. However, the responsiveness of the F-actin-network has an
optimal value as in the full model.

Further simulations after successive variation of the frozen
parameters (see above) reveal, that an optimal range of adhesiveness
could be gained only if a passive frictional load $\Phi_u>0$ was
chosen to increase monotonically with adhesion strength
$\mathrm{Adh0}$, as it was assumed in the full model, see
\begin{table}[htbp]
\centering 
\footnotesize{
\begin{tabular}{llll} 
\textbf{Symbol} & \textbf{Meaning} & \textbf{Value} & \textbf{Units}
\\ \hline $a_g$            &  Concentration of G-actin
                         & $30$   & $\mu \mathrm M$ \\
$a_\mathrm{max}$ &  Maximal concentration of F-actin
                         & $800$  & $\mu \mathrm M$ \\
$k_\mathrm{on}$  &  F-actin polymerization rate at plus ends
              & $696$ & $(\mathrm{min} \cdot \mu \mathrm M)^{-1}$ \\
$k_\mathrm{off}$ &  F-actin depolymerization rate at plus ends
                         & $258$  & $\mathrm{min}^{-1}$ \\
$\omega$      & F-actin capping rate          & $1250$ & $\mathrm{min}^{-1}$ \\
$r$              &  F-actin disassembly rate
                         & $0.2$ & $\mathrm{min}^{-1}$ \\%
$\varepsilon$ & Basal F-actin nucleation rate & $0.75$
                             & $\mathrm{min}^{-1}$ \\
Arp0          &  Concentration of activated Arp2/3 complexes
                         & $10$  & $\mu \mathrm M$ \\
$\nu_a$       & Arp2/3-induced nucleation rate
                      & $60$         & $(\mathrm{min} \cdot \mu \mathrm M)^{-1}$ \\
$K_a$         & Half-saturation concentration for nucleation
                      & $3$          & $\mu \mathrm M$ \\
$m_f^0$       & Equilibrium reservoir concentration of free
myosin-II
                      & $10$         & $\mu \mathrm M$ \\
$D_m$         & Diffusion coefficient for free myosin-II
                      & $0.5$        & $\mu \mathrm m^2 \cdot \mathrm{min^{-1}}$ \\
$\alpha_m$    & F-actin binding rate of myosin-II
                      & $0.5$  & $(\mathrm{min} \cdot \mu \mathrm M)^{-1}$ \\
$\delta_m^0$  & Dissociation rate of bound myosin-II
                      & $3$          & $\mathrm{min}^{-1}$ \\
$a_{opt}$     & Optimal F-actin concentration for myosin-II binding
                      & $40$         & $\mu \mathrm M$ \\
$c_f^0$       & Reservoir concentration of free integrin at lamellar
tips
                      & $50$         & $\mu \mathrm m^{-2}$ \\
$D_f$         & Diffusion coefficient for free integrin
                      & $0.5 $       & $\mu \mathrm m^2 \cdot \mathrm{min^{-1}}$ \\
$\mathrm{Adh0}$ & Number of available substrate sites per integrin  & $3$          & dimensionless \\
Talin         & F-actin association factor         & $0.0125$ & dimensionless \\
Fifactor      & Focal complex factor (intracellular) & $50$    &  dimensionless \\
Fefactor      & Focal adhesion factor (extracellular) & $50$     & dimensionless \\
$\alpha^-$    & Dissociation rate for substrate bound integrin
                      & $\alpha = 5$ & $\mathrm{min}^{-1}$ \\
$\alpha^+$    & Free integrin binding rate to substrate
                      & $\mathrm{Adh0}\cdot \alpha$ & $\mathrm{min}^{-1}$ \\
$\beta^+_0$   & F-actin binding rate for substrate bound integrin
                      & $\alpha \cdot \text{Talin}$
                      & $(\mathrm{min} \cdot \mu \mathrm M)^{-1}$ \\
$\beta^-_0$   & FA dissociation of the F-actin link
                      & $\alpha/\text{Fifactor}$ & $\mathrm{min}^{-1}$ \\
$\delta^+_0$  & Free integrin binding rate to F-actin
                      & $\alpha\cdot \text{Talin}$
                      & $(\mathrm{min} \cdot \mu \mathrm M)^{-1}$ \\
$\delta^-$    & Dissociation rate for F-actin bound integrin
                      & $\alpha$     & $\mathrm{min}^{-1}$ \\
$\gamma^+$    & Substrate binding rate for F-actin bound integrin
                      & $\mathrm{Adh0}\cdot \alpha $
                      & $\mathrm{min}^{-1}$ \\
$\gamma^-_0$  & FA dissociation rate of the substrate link
                      & $\alpha/\text{Fefactor}$ & $\mathrm{min}^{-1}$ \\
$\rho_\beta$  & Exponential FA-rupture coefficient for F-actin link
                      & $11.7$       & $(\mathrm{Pa}\cdot\mu \mathrm m)^{-1}$ \\
$\rho_\gamma$ & Exponential FA-rupture coefficient for substrate
link
                      & $11.7$       & $(\mathrm{Pa}\cdot\mu \mathrm m)^{-1}$ \\
$A$           & Total free F-actin binding sites on boundary
$\Gamma(t)$
                      & $50$         & $\mu \mathrm M $ \\
$\pi_0^f$     & Ratchet polymerization pressure coefficient
                      & $0.0125$     & $\mathrm{Pa} \cdot\mu \mathrm M^{-1}$ \\
$\pi_0^c$     & Clamp-motor polymerization pressure coefficient
                      & $0.0125$     & $\mathrm{Pa} \cdot\mu \mathrm M^{-1}$ \\
$\kappa_\Gamma$ & Tip curvature weight factor for membrane tension
                      & $0.5$     & dimensionless \\
$\alpha_0^f$ & Relative amount of exposed barbed ends per Arp2/3
                      & $2$       & dimensionless \\
$\psi_0$      & Contractile stress per bound myosin-II
                      & $0.0163$   & $\mathrm{Pa}\cdot \mu \mathrm M^{-1}$ \\
$\widetilde{\psi}_0$ & Strength of contractile stress (simplified
model)
                      & $0.625$   & $\mathrm{Pa}$ \\
$\sigma_0$    & Strength of the swelling pressure
                      & $0.125$      & $\mathrm{Pa}$ \\
$\mu$         & Viscosity of the F-actin network phase
                      & $0.625$      & $\mathrm{Pa} \cdot \mathrm{min}$ \\
$\kappa$      & Cortical slip parameter for the F-actin flow
                      & $0.5$        & dimensionless \\
$\varphi$        & Drag coefficient between network and solvent
                      & $2$
                      & $\mathrm{Pa} \cdot \mathrm{min} \cdot \mu \mathrm m^{-2}$ \\
$\Phi_0$      & Friction per actin-substrate bound integrin
                      & $0.02$
                      & $\mathrm{Pa} \cdot \mathrm{min} $ \\
$\Phi_u$      & Additional friction associated to the cell body
                      & $\mathrm{Adh0}\cdot 6$
                      & $\mathrm{Pa} \cdot \mathrm{min} \cdot \mu \mathrm m^{-2}$ \\
$L$           & Length of the cell (fragment)      & $10$   & $\mu \mathrm m$ \\
\hline
\end{tabular}
} \caption{List of parameters for the 1-D simulations, if not
otherwise specified.}\label{tab:parameter}
\end{table}
Table~\ref{tab:parameter} and Fig.~\ref{fig:CurvesMax}(a). Though
such a friction appears to be physically realistic, the question
remains whether such an optimal velocity response is of general
biological relevance. For, the response of living systems is
typically guided by adaptation: Migrating cells show the capability
to dissolve their focal adhesions by proteolytic cleavage of
integrins \cite{satish05} which could be an adaptive strategy to
optimize cellular adhesiveness on various substrates. On the other
hand, the adhesiveness could be raised by active segregation of
extracellular matrix proteins.

Nevertheless, the universal optimality curve in dependence of
F-actin responsiveness, seen in both Figs.~\ref{fig:CurvesMax}(b)
and \ref{fig:integrins5}(b), seems to hold for any mechanical model,
which implements a saturating F-actin polymerization by association
of regulating proteins as Arp2/3. This indicates that an optimal
control of migration velocity by tuning certain chemical processes
as F-actin polymerization, is more easy to be realized than tuning
adhesion, and thus may have evolved as a generally effective
strategy to regulate such mechanical processes as cell adhesion and
locomotion.


\section{Discussion and outlook to further modeling}

Based on a larger set of interwoven mass and force balance equations
for mechano-chemical processes within the cytoplasm and the
surrounding plasma membrane, the presented continuum model is a
simplified though already sufficiently complex version of a more
comprehensive 3-dimensional whole-cell model. From the 1-D and 2-D
simulation results we conclude that even without organizing centers
(as the cell nucleus or microtubuli, cf.~\cite{Kaverina02}) or
regulating systems (as the Rho/Rac control cascades,
cf.~\cite{ridley:integrating}) the F-actin cytoskeleton and its
associated proteins having mechanical functions (myosin and
integrin) constitute a self-organizing biophysical system with the
ability of autonomous polarization and locomotion. The transition
form a symmetric, unpolarized stationary state into a polarized
migrating state can be mechanically induced or spontaneous due to
stochastic or chaotic fluctuations, depending on the
(meta-)stability of the stationary state. Clearly, for reproducing
the mentioned experimental data, the 2-dimensional system offers a
wider spectrum of possibilities, one of which has been elaborated by
Kozlov and Mogilner \cite{mogilner:polarization}. They prove a
defined bi-stability between the radially symmetric cell shape and a
polarized, circularly indented shape by allowing for different
anisotropic organization of the actin-myosin cytoskeleton. In
comparison to this, further numerical `experiments' using our
coupled viscous flow-transport-diffusion-reaction system with free
boundary should be performed to explore the capability of isotropic
continuum models. Though local orientation or alignment of actin
filaments will have an enforcing effect onto the dynamics,
interactions and feedback between F-actin, cross-linkers,
motor-proteins and adhesion complexes (for a 1-dimensional model see
\cite{stevens:symmetrybreaking}), already the presented simulation
results show that internal concentration gradients and directional
flow, leading to persistent polarization, can emerge from local
fluctuations. Thereby super-threshold pattern formation arises, not
as in common excitable media by a purely reaction-diffusion
feedback, rather by a spatially distributed coupling between
chemically induced force generation/relaxation and mechanically
induced bond formation/dissociation.

More than one decade ago, Bereiter-Hahn and L\"uers
\cite{bereiter:subcellular} claimed that, based on their experiments
and measurements with migrating keratocytes, polarization and
directionality of cells are determined by local variations in actin
network stiffness and hydrostatic pressure. Together with the more
recent high-resolution measurements of dynamic vector field (as
F-actin velocity, forces onto pliable substrates, or directionality
of FA shape deformations), it could soon be possible to probe and
evaluate the various assumptions and hypotheses on mass and force
balance conditions, which we postulate in
Sections~\ref{sec:massbalanceeqs} and \ref{sec:forcebalanceeqs}. For
example, the biphasic correlations between retrograde actin flow
speed and traction stress detected by Gardel~et~al.~\cite{Gardel08}
could well be resolved by the measured and simulated FA gradient
away from an advancing lamellar tip, and by our central modeling
assumption that the local traction force $\mathbf F_v$ is
proportional to the product $c_{sa}\cdot\mathbf v_c$ of FA
concentration and cortical F-actin velocity.

Moreover, the distribution of flow and force, extensively discussed
for the 1-dimensional case (in Section~\ref{sec:induced}), has to be
quantitatively characterized for the 2-dimensional free boundary
situation (Section~\ref{sec:spontan}), where the overall picture is
the same: Wider distribution of traction forces in the leading front
region (with maximum flow speed at the tip, depending on the active
polymerization pressure) and strong disruptive counter-forces in a
confined region of `rear release'.
The actin flow patterns of our 2-D simulations show in principle the
same characteristics as experimental data by Yam et
al.~\cite{pmid17893245} 
for fish keratocytes. During the process of polarization,
the simulated cell looses its radial symmetry of a uniform inward
flow pattern, leading to a stronger inward flow in the rear region,
whereby the most pronounced forces come from the two sides relative
to the establishing locomotion direction, see
Fig.~\ref{fig:cellFragment}(c).
Indeed, traction force experiments on polarized fish keratocytes
have revealed the same pattern, with major traction forces from
the two flanks at the rear and weaker retrograde forces at the front
\cite{oliverdembo}.

Further evaluation of the presented model should compare the simulated
integrin concentration profiles with more precise quantitative data
on the temporal turnover and spatial distribution of FAs in
migrating cells, as already mentioned above. For this reason, the
model has to be extended to include stochastic clustering dynamics
of integrins, which probably is an essential feature of FAs
determining the detailed patterns of
F-actin flow and traction forces.
We emphasize, that our model
results do not depend on the specific mechanisms of myosin-II
contraction. Essential for the emergence of coupled flow and mass
gradients is the bimodal (cubic-like) stress function $S =
S(\theta)$ in Fig.~\ref{fig:assembly} which similarly appears in a
modified mechanical situations as, for example, the nematode sperm
motility system, see \cite{bottino:nematode}.

The experimental response curves for the migration speed of
polarized cells depending on adhesiveness \cite{Palecek97} show a
broad optimum range with a logarithmic slow decay for larger Adh0
values and a steep decay for lower ones, similar to the curves found
in our 1-D model, Fig.~\ref{fig:CurvesMax}, and in earlier analogous
simulation results using alternative visco-elastic mechanics
\cite{DiMillaLauffenburger,grachevaothmer}. The deviating behavior
for very low adhesiveness values hat still to be investigated with a
more precise quantification of other (weaker) substrate-mediated
forces that may come into account.

An optimum curve for the protrusion speed in dependence of the
barbed end density $B$, thus also of the concentration Arp0, has
been obtained by Mogilner and Edelstein-Keshet \cite{Mog02} with the
aid of 1-dimensional diffusion-reaction-transport systems, including
membrane resistance and the brownian ratchet mechanism. However,
they do not model retrograde actin flow or force transduction to the
substratum, so that the shape of their curve (\cite{Mog02},Fig. 6)
essentially differs from our results in Figs.~\ref{fig:CurvesMax}(b)
and \ref{fig:CurvesFcsa}(b).

Finally, in generalization of the model simulations by
St\'ephanou~et~al.~\cite{StephanouTracqui} and in comparison to
other modeling approaches mentioned before, our full 2-dimensional
model should be used for reproducing the experimentally observed
translocation paths and deformation structures of single blood or
tissue cells in culture, which show well-expressed phases of
persistent polarization and locomotion, interrupted by events of
speed reduction, contractile rounding, successive re-polarization
and migration. Statistical analysis, in the same spirit as having
been performed for the original annual lamella model
\cite{AltTranquillo95}, could help to reveal the microscopic
mechanisms that are responsible for cell motility behavior on the
macroscopic level.

\paragraph{Acknowledgments:}
This joint work has been generously supported by a grant on
"Simulation models for cell motility -- coupling substrate adhesion 
and cytoplasm dynamics" (I/80543,
Volkswagen-Stiftung). 
Moreover, the hospitality of the Soft Matter
Physics group at Leipzig University during the final preparation of this
chapter is highly appreciated.
Finally we thank our colleagues at Bonn
University and the Research Center J\"ulich for helpful discussions
and critical remarks.

  \bibliographystyle{plain}
  \bibliography{alt_bock_moehl}
\end{document}